\documentclass[english, oneside, fontsize=11pt, final, a4paper]{scrartcl}

\pdfoutput=1

\usepackage[T1]{fontenc}
\usepackage[utf8]{inputenc}
\usepackage{babel}

\usepackage{nccmath} % fleqn environment
\usepackage{mathtools} % loads and fixes amsmath
\usepackage{amsfonts}
\usepackage{amssymb}
\usepackage{dsfont}
\usepackage{physics}
\usepackage{slashed}
\numberwithin{equation}{section}

\usepackage{subcaption}

\usepackage[final, babel=true]{microtype}
\usepackage{csquotes} % automatic quotation marks, also used for bibliography
\usepackage{graphicx}
\usepackage{booktabs} % pretty tables

\makeatletter
\g@addto@macro\bfseries{\boldmath}
\makeatother

\usepackage[nosort]{cite}

\usepackage[table]{xcolor}
\usepackage{hyperref}
\usepackage[ocgcolorlinks]{ocgx2} % have link colors black when printing
	\hypersetup{
		final,
		unicode=true,
		allcolors={linkcolor},
		pdftitle={On quantum deformations of AdS₃⨉S³⨉T⁴ and mirror duality},
		pdfauthor={Fiona K. Seibold, Stijn J. van Tongeren, and Yannik Zimmermann},
		pdfdisplaydoctitle=true,
	}
\usepackage{bookmark}
\usepackage{cleveref}

\usepackage{arydshln}

\newcommand{\resizeToFitPageMath}[1]{\resizebox{\textwidth}{!}{$#1$}}

\newcommand{\numberthis}{\stepcounter{equation}\tag{\theequation}}

\newcommand{\mup}[1]{{\textnormal{#1}}}
\newcommand{\mdot}{\:.}
\newcommand{\mcomma}{\:,}
\newcommand{\I}{i} % imaginary unit
\newcommand{\E}{e} % Euler number

\DeclareMathOperator{\arsinh}{arsinh}

\newcommand{\AdS}{\mup{AdS}}
\newcommand{\adsThree}{\textup{AdS}_\textup{3}\times\textup{S}^\textup{3}}
\newcommand{\adsThreeT}{\textup{AdS}_\textup{3}\times\textup{S}^\textup{3}\times\textup{T}^\textup{4}}
\newcommand{\adsFive}{\textup{AdS}_\textup{5}\times\textup{S}^\textup{5}}

\newcommand{\dsS}{\mathbb{S}}
\newcommand{\dsT}{\mathbb{T}}
\newcommand{\scS}{\mathcal{S}}
\newcommand{\scT}{\mathcal{T}}
\newcommand{\scR}{\mathcal{R}}
\newcommand{\identity}{\mathds{1}}
\newcommand{\conjugated}[1]{#1^*}

\newcommand{\defoFormat}[1]{\texttt{#1}}
\newcommand{\fermOne}{\defoFormat{ferm1}}
\newcommand{\fermTwo}{\defoFormat{ferm2}}
\newcommand{\distOne}{\defoFormat{dist1}}
\newcommand{\distTwo}{\defoFormat{dist2}}
\newcommand{\xoxOne}{\defoFormat{xox1}}
\newcommand{\xoxTwo}{\defoFormat{xox2}}
\newcommand{\xox}{xox}

\newcommand{\massPhasezeta}{\beta_\zeta}
\newcommand{\massPhasechi}{\beta_\chi}
\newcommand{\fermionPhase}{\beta}
\newcommand{\omegaNormalization}{\tilde{\omega}}
\newcommand{\operatorPhase}{\varphi}
\newcommand{\operatorPhaseYPlus}{\operatorPhase_{Y_+}}
\newcommand{\operatorPhaseYMinus}{\operatorPhase_{Y_-}}
\newcommand{\operatorPhaseYPlusMinus}{\operatorPhase_{Y_\pm}}
\newcommand{\operatorPhaseZPlus}{\operatorPhase_{Z_+}}
\newcommand{\operatorPhaseZMinus}{\operatorPhase_{Z_-}}
\newcommand{\operatorPhaseZPlusMinus}{\operatorPhase_{Z_\pm}}
\newcommand{\operatorPhaseZetaPlus}{\operatorPhase_{\zeta_+}}
\newcommand{\operatorPhaseZetaMinus}{\operatorPhase_{\zeta_-}}
\newcommand{\operatorPhaseZetaPlusMinus}{\operatorPhase_{\zeta_\pm}}
\newcommand{\operatorPhaseChiPlus}{\operatorPhase_{\chi_+}}
\newcommand{\operatorPhaseChiMinus}{\operatorPhase_{\chi_-}}
\newcommand{\operatorPhaseChiPlusMinus}{\operatorPhase_{\chi_\pm}}

\def\wasyfamily{\fontencoding{U}\fontfamily{wasy}\selectfont}
\def\Circle{\mbox{\wasyfamily\char35}}

\newcommand{\stringtension}{h} %to be able to easily adapt naming later
\newcommand{\supertranspose}{{\scriptscriptstyle \textsf{ST}}}

\definecolor{linkcolor}{HTML}{00445C} % blue

\definecolor{color1}{HTML}{A9E6D5} % green
\definecolor{color2}{HTML}{F7E8C3} % yellow
\definecolor{color3}{HTML}{B3E4FF} % blue
\definecolor{color4}{HTML}{FFE7D3} % red

\addtokomafont{titlehead}{\raggedleft\ttfamily}
\titlehead{Imperial-TP-FS-2021-01\\HU-EP-21/19}

\title{On quantum deformations of $\adsThreeT$ and mirror duality}
\date{}

\newcommand*{\affaddr}[1]{\normalsize\textit{#1}}
\newcommand*{\affmark}[1]{\textsuperscript{#1}}
\newcommand*{\affmarkpunc}[1]{\hspace{-0.1em}\affmark{#1}}
\newcommand*{\email}[1]{\normalsize\texttt{#1}}

\author{
	Fiona K. Seibold,\affmarkpunc{a,b}
	Stijn J. van Tongeren,\affmarkpunc{c}
	and
	Yannik Zimmermann\affmark{c}
	\vspace{1cm} \\
	\affmark{a}\affaddr{Institut für Theoretische Physik, ETH Zürich,}\\
	\affaddr{Wolfgang-Pauli-Strasse 27, 8093 Zürich, Switzerland}
	\vspace{0.5cm} \\
	\affmark{b}\affaddr{Blackett Laboratory, Imperial College London,}\\
	\affaddr{London SW7 2AZ,  United Kingdom}
	\vspace{0.5cm} \\
	\affmark{c}\affaddr{Institut für Physik, Humboldt-Universität zu Berlin,}\\
	\affaddr{IRIS Gebäude, Zum Grossen Windkanal 6, 12489 Berlin, Germany}
	\vspace{1cm} \\
	\email{fseibold@itp.phys.ethz.ch} \\
	\email{svantongeren@physik.hu-berlin.de} \\
	\email{yannik.zimmermann@physik.hu-berlin.de}
}

\begin{document}

\maketitle

\begin{abstract}
\noindent
We consider various integrable two-parameter deformations of the $\adsThreeT$ superstring with quantum group symmetry. Working on the string worldsheet in light-cone gauge and to quadratic order in fermions, we obtain their common massive tree-level two-body S matrix, which matches the expansion of the conjectured exact $q$-deformed S matrix. We then analyze the behavior of the exact S matrix under mirror transformation -- a double Wick rotation on the worldsheet -- and find that it satisfies a mirror duality relation analogous to the distinguished $q$-deformed $\adsFive$ S matrix in the one parameter deformation limit. Finally, we show that the fermionic $q$-deformed $\adsFive$ S matrix also satisfies such a relation.
\end{abstract}

\pagebreak
\pdfbookmark[section]{\contentsname}{toc} % add toc to pdf bookmarks
\tableofcontents

\section{Introduction}

The presence of integrable structures in various instances of the AdS/CFT correspondence has led to remarkable insight into both gauge and string theory \cite{Beisert:2010jr,Bombardelli:2016rwb}. This motivates the search for integrable deformations of these models, which has been fruitfully pursued in particular in string theory, in the form of Yang-Baxter deformed sigma models \cite{Klimcik:2002zj,Klimcik:2008eq,Delduc:2013qra}. There is a variety of Yang-Baxter deformations, with distinct algebraic properties and interpretations in terms of string theory and AdS/CFT. In this paper we focus on inhomogeneous Yang-Baxter deformations -- algebraically leading to trigonometric quantum (q) deformed algebras  -- also referred to as $\eta$ deformations. These deformations are determined by a so-called $R$ operator, which solves the modified (inhomogeneous) classical Yang-Baxter equation (mCYBE).\footnote{There are also homogeneous Yang-Baxter deformations \cite{Kawaguchi:2014qwa}, including e.g.\ the well-known real $\beta$ deformation of the $\adsFive$ string \cite{Matsumoto:2014nra}, which algebraically correspond to twisted symmetry \cite{vanTongeren:2015uha,vanTongeren:2018vpb}, see also \cite{Kawaguchi:2013lba}. This twisted symmetry has been used to conjecture field theory duals \cite{vanTongeren:2015uha}.}

While Yang-Baxter deformations preserve kappa symmetry \cite{Delduc:2013qra}, this is not sufficient to guarantee that the corresponding backgrounds satisfy the supergravity equations of motion \cite{Arutyunov:2015qva}. In general these backgrounds only satisfy a set of generalized supergravity equations \cite{Arutyunov:2015mqj}, which derive from kappa symmetry \cite{Wulff:2016tju}. These generalized equations should guarantee scale invariance, but not Weyl invariance \cite{Arutyunov:2015mqj,Wulff:2016tju,Wulff:2018aku}.\footnote{There have been further proposals regarding Weyl invariance for generalized supergravity backgrounds \cite{Fernandez-Melgarejo:2018wpg,Muck:2019pwj}. These proposals have certain troublesome features, however, as discussed in \cite{Muck:2019pwj}.} A sufficient condition for Weyl-invariance is unimodularity of the $R$ operator \cite{Borsato:2016ose,Hronek:2020skb}. For superalgebras, the freedom to chose different Dynkin diagrams underlying the canonical Drinfel'd-Jimbo solution to the mCYBE, allows one to find unimodular inhomogeneous deformations \cite{Hoare:2018ngg,Seibold:2019dvf}. Starting from fermionic Dynkin diagrams give unimodular supergravity backgrounds, while others yield only solutions of the generalized supergravity equations.

The original Yang-Baxter deformation of $\adsFive$ \cite{Delduc:2013qra,Arutyunov:2015qva} is based on the distinguished Dynkin diagram, and does not correspond to a supergravity background. Previous studies of its integrable structure show a particularly interesting feature however. At the level of the spectrum and exact S matrix \cite{Beisert:2008tw,Hoare:2011wr}, this distinguished $\eta$ deformation of $\adsFive$ displays so-called \enquote{mirror duality} \cite{Arutynov:2014ota,Arutyunov:2014cra,Arutyunov:2014jfa,Pachol:2015mfa}. In this class of models, combining the mirror transformation -- a double Wick rotation interchanging worldsheet space and time -- with inversion of the deformation parameter, $\kappa\rightarrow 1/\kappa$, is a symmetry of the scattering theory. This interestingly relates the spectral and thermodynamic properties of pairs of models with relatively inverse deformation parameters. Independently, it is possible to consider the effect of the mirror transformation, at the level of the gauge fixed string action, and translate this to a set of transformation rules for the background fields of the sigma model \cite{Arutyunov:2014cra,Arutyunov:2014jfa}. This gives what we can call a geometric mirror transformation. We can then ask if the ($\kappa$-dependent) background fields are compatible with mirror duality. At the level of the metric and $B$ field this is the case, however the Ramond-Ramond (RR) forms of the distinguished deformation are not manifestly compatible with this structure. In fact the mirror transformation here gives a complex background. This leads to the question whether there is some other inhomogeneous deformation of $\adsFive$ which does have manifest mirror duality at the geometric level. For $\adsFive$ the answer appears to be negative. In particular, also the RR forms of the unimodular fermionic deformation of $\adsFive$ break geometric mirror duality \cite{Hoare:2018ngg}, while mirror duality at the spectral level has thus far not been investigated in this case. Interestingly, applying the geometric mirror transformation to \emph{undeformed} $\adsFive$ yields a supergravity background, mirror $\adsFive$. It is an open question to understand whether this integrable model can be related to the maximal deformation limit -- mirroring the undeformed limit -- of some inhomogeneous Yang-Baxter deformation of $\adsFive$.

Moving beyond $\adsFive$, deformations of $\adsThreeT$ offer an interesting area of investigation, due to extra freedom with interesting consequences. Firstly, the group structure of AdS$_3$ and S$^3$ makes it possible to define two \cite{Klimcik:2014bta,Hoare:2014oua,Seibold:2019dvf} and three-parameter deformations \cite{Delduc:2018xug} of this background. At least in the two-parameter case some of these are also known to correspond to supergravity backgrounds \cite{Seibold:2019dvf}. Secondly, in contrast to $\adsFive$, here it is possible to find a (single-parameter) unimodular inhomogeneous deformation which is manifestly invariant under mirror duality at the geometric level \cite{Seibold:2019dvf}.

In this paper, we compute the massive tree level S matrices for various two-parameter deformations of $\adsThreeT$, including fermions to quadratic level (sections \ref{sec:interaction_Lagrangian} and \ref{sec:Smat_perturbative}), and contrast these with the conjectured exact S matrix \cite{Hoare:2014oua} (section \ref{sec:Smat_exact}).\footnote{Determining the exact dressing phase(s) for two-parameter deformations is an open problem. In general, determining the complete $\adsThreeT$ exact S matrix, including massless excitations and dressing phases, is an involved problem already in the undeformed case, see e.g. \cite{Borsato:2016xns}.} We consider all deformations of $\adsThreeT$ which use the same Dynkin diagram for the $R$ operator in each factor of the symmetry algebra $\mathfrak{psu}(1,1|2)^{\oplus2}$. Two of them are based on the fermionic Dynkin diagram, are hence unimodular, with one giving manifestly mirror dual backgrounds for one-parameter deformations. The other four cases are not unimodular, two based on the distinguished Dynkin diagram, and two on the \xox{} one. Despite these backgrounds being geometrically distinct, the minimal rank of the light-cone symmetry algebra means that there is only one Dynkin diagram and hence essentially a unique exact S matrix common to all these models. This is in interesting contrast to $\adsFive$, where the fermionic and distinguished deformations have inequivalent S matrices \cite{Seibold:2020ywq}. For $\adsThreeT$, different inhomogeneous deformations manifest themselves as one particle changes of basis.\footnote{Similar observations were made regarding the effect of different choices of real form underlying inhomogeneous deformations, in \cite{Hoare:2016ibq}.} Our results give an overview of the perturbative structure of the single particle phases relating the ``different'' S matrices. Our results match the two-parameter limit of the tree-level bosonic S matrix of the three-parameter deformations of $\adsThreeT$ computed in \cite{Bocconcello:2020qkt}.\footnote{Further related work on a subsector of the three-parameter deformed bosonic S matrix recently appeared in \cite{Garcia:2021iox}.}

Beyond our tree-level verification of the exact S matrix, we also revisit the topic of mirror duality. First, in section \ref{sec:mirror_duality}, we show that the exact S matrix for single-parameter deformations of $\adsThreeT$ satisfies a mirror-duality relation analogous to one of the distinguished $q$-deformed $\adsFive$ S matrix. This guarantees mirror duality at the level of the spectrum, modulo certain assumptions on the as of yet not fully known dressing phases. This relationship holds for all deformations of $\adsThreeT$ which we considered. We also discuss why it is unlikely that the two-parameter deformed S matrix could have mirror duality. Second, in section \ref{sec:mirror_duality_AdS5}, we come back to the question of mirror duality for the fermionic deformation of $\adsFive$. Because the S matrix for this deformation differs from the distinguished one, and the fermionic deformed background is not manifestly mirror self dual, it was a priori not clear whether the spectrum of this Weyl-invariant q deformation of $\adsFive$ has mirror duality. We verify that the fermionic deformed S matrix satisfies the same mirror duality relations as the distinguished deformed S matrix, manifesting mirror duality at the spectral level also here.

\section{Determining the interaction Lagrangians}
\label{sec:interaction_Lagrangian}

We would like to compute the tree-level two-body worldsheet S matrix of the massive excitations of various inhomogeneous deformations of the $\mathrm{AdS}_3\times \mathrm{S}^3\times \mathrm{T}^4$ string in the light-cone gauge. To do so, we need the corresponding action for each deformation in the light-cone gauge, expanded to quartic order in the fields. In our previous paper \cite{Seibold:2020ywq} we determined this expanded action for deformations of $\mathrm{AdS}_5\times \mathrm{S}^5$. Here we follow the same approach and conventions, applied to various deformations of $\mathrm{AdS}_3\times \mathrm{S}^3 \times \mathrm{T}^4$.

\subsection{\texorpdfstring{Two-parameter deformations of $\adsThreeT$}{Two-parameter deformations of AdS₃⨉S³⨉T⁴}}

The two-parameter inhomogeneous deformations of $\adsThreeT$ that we are considering, share the metric and $B$ field \cite{Hoare:2014oua}
\begin{equation}\label{eq:metricandBfield}
\begin{aligned}
\dd s^2 = & \frac{1}{F(\rho)} \left[-(1+\rho^2) \dd t^2 + \frac{\dd \rho^2}{1+\rho^2} + \rho^2 \dd \psi^2 - (\kappa_- (1+\rho^2) \dd t - \kappa_+ \rho^2 \dd \psi)^2 \right] \\
&+ \frac{1}{\tilde{F}(r)} \left[(1-r^2) \dd \varphi^2 + \frac{\dd r^2}{1-r^2} + r^2 \dd \phi^2 + (\kappa_-(1-r^2) \dd \varphi +  \kappa_+ r^2 \dd \phi)^2 \right] \\
& \phantom{+} + \dd x^i \dd x^i~,\\
B &=\frac{\rho}{F(\rho)} (\kappa_+ \dd t \wedge \dd \rho + \kappa_- \dd \rho \wedge \dd \psi)+ \frac{ r}{\tilde{F}(r)} ( \kappa_+  \dd \varphi \wedge \dd r + \kappa_- \dd r \wedge \dd \phi)~,
\end{aligned}
\end{equation}
where the $x^i$, $i=1,\dots,4$ are the torus coordinates, $\kappa_\pm$ are the two deformation parameters, and
\begin{equation}
F(\rho)= 1+\kappa_-^2(1+\rho^2)-\kappa_+^2\rho^2~, \qquad \tilde{F}(r)= 1+\kappa_-^2(1-r^2)+\kappa_+^2r^2~.
\end{equation}

Including fermions, we will consider the backgrounds constructed in \cite{Seibold:2019dvf}, namely two deformations based on the fermionic Dynkin diagram of $\mathfrak{psu}(1,1|2)$ (denoted \fermOne{} and \fermTwo{}), two based on the distinguished Dynkin diagram (\distOne{} and \distTwo{}), and finally two based on the \xox{} one (\xoxOne{} and \xoxTwo{}). Only the backgrounds based on the fermionic Dynkin diagram correspond to solutions of the supergravity equations of motion. The first of these, \fermOne{}, has dilaton $\Phi$ and RR fluxes $F=dC$ with
\begin{align}\label{eq:fermionicbackground1}
e^{-2 \Phi} &= e^{-2 \Phi_0} \frac{F(\rho)\tilde{F}(r)}{P(\rho,r)^2}~, \\
C_2 &= - \sqrt{\frac{1+\kappa_+^2}{1+\kappa_-^2}} \frac{e^{-\Phi_0}}{P(\rho,r)} \big[ \rho^2 \dd t \wedge \dd \psi + r^2 \dd \varphi \wedge \dd \phi  + \kappa_-^2(1+\rho^2) r^2 \dd t \wedge \dd \phi \nonumber\\
&\qquad- \kappa_-^2 \rho^2 (1-r^2) \dd \psi \wedge \dd \varphi + \kappa_+ \kappa_- (\rho^2-r^2-\rho^2 r^2) \dd t \wedge \dd \varphi - \kappa_+ \kappa_- \rho^2 r^2 \dd \psi \wedge \dd \phi \big]~,\nonumber\\
C_4 &= - \sqrt{\frac{1+\kappa_+^2}{1+\kappa_-^2}} \frac{e^{-\Phi_0}}{P(\rho,r)} \big[ \kappa_- \rho^2 \dd t \wedge \dd \psi + \kappa_- r^2 \dd \varphi \wedge \dd \phi - \kappa_- (1+\rho^2) r^2 \dd t \wedge \dd \phi \nonumber\\
&\qquad+ \kappa_- \rho^2 (1-r^2) \dd \psi \wedge \dd \varphi - \kappa_+ (\rho^2-r^2-\rho^2 r^2) \dd t \wedge \dd \varphi + \kappa_+ \rho^2 r^2 \dd \psi \wedge \dd \phi \Big] \wedge J_2~,\nonumber
\end{align}
where
\begin{equation}
\label{eq:torusKahlerform}
J_2= \dd x^1 \wedge \dd x^2 - \dd x^3 \wedge \dd x^4,
\end{equation}
and
\begin{equation}
P(\rho,r) = 1-\kappa_+^2 (\rho^2-r^2-\rho^2 r^2)+\kappa_-^2(1+\rho^2)(1-r^2)~.
\end{equation}
The \fermTwo{} background corresponds to
\begin{align}\label{eq:fermionicbackground2}
e^{-2 \Phi} &= e^{-2 \Phi_0} \frac{F(\rho)\tilde{F}(r)}{Q(\rho,r)^2}~, \\
C_2 &= - \sqrt{\frac{1+\kappa_-^2}{1+\kappa_+^2}} \frac{e^{-\Phi_0}}{Q(\rho,r)} \big[ (1+\rho^2) \dd t \wedge \dd \psi -(1-r^2) \dd \varphi \wedge \dd \phi  + \kappa_+^2 (1+\rho^2) r^2 \dd t \wedge \dd \phi \nonumber\\
&\qquad- \kappa_+^2 \rho^2(1-r^2) \dd \psi \wedge \dd \varphi + \kappa_+ \kappa_- (1+\rho^2) (1-r^2) \dd t \wedge \dd \varphi - \kappa_+ \kappa_- (1+\rho^2 r^2) \dd \psi \wedge \dd \phi  \big]~,\nonumber\\
C_4 &= - \sqrt{\frac{1+\kappa_-^2}{1+\kappa_+^2}} \frac{e^{-\Phi_0}}{Q(\rho,r)} \big[ \kappa_+  (1+\rho^2) \dd t \wedge \dd \psi - \kappa_+ (1-r^2) \dd \varphi \wedge \dd \phi - \kappa_+ (1+\rho^2) r^2 \dd t \wedge \dd \phi \nonumber\\
&\qquad+ \kappa_+ \rho^2 (1-r^2) \dd \psi \wedge \dd \varphi - \kappa_-(1+\rho^2) (1-r^2) \dd t \wedge \dd \varphi  + \kappa_- (1+\rho^2 r^2) \dd \psi \wedge \dd \phi  \Big] \wedge J_2~,\nonumber
\end{align}
with
\begin{equation}
Q(\rho,r) = 1-\kappa_+^2 \rho^2 r^2+\kappa_-^2(1+\rho^2 r^2)~.
\end{equation}
The second of these backgrounds is manifestly compatible with mirror duality for the one-parameter deformation $\kappa_+ = \kappa$, $\kappa_- = 0$. The remaining (generalized supergravity) backgrounds we consider are given in appendix \ref{app:generaliedsugrabackgrounds}.

\subsection{Gauge fixing and expansion}

Given the above backgrounds, we proceed analogously to our previous discussion for deformations of $\mathrm{AdS}_5\times \mathrm{S}^5$ \cite{Seibold:2020ywq}, where all technical details can be found. Here we briefly indicate the adaptations required for deformations of $\mathrm{AdS}_3 \times \mathrm{S}^3 \times \mathrm{T}^4$. The coordinates $t$ and $\varphi$ are used to form the light-cone coordinates $x^\pm$, and we change our basis of transverse fields from $\rho, \psi, r, \phi$ to $z_1,z_2,y_1,y_2$, via
\begin{equation}\label{eq:zyparametrization}
\rho \, e^{ i \psi}  = \frac{z_1 + i z_2}{1-z^2/4}, \qquad r\, e^{i \phi} = \frac{y_1 + i y_2}{1+y^2/4}
\mcomma
\end{equation}
where $z^2=z_1^2+z_2^2$ and $y^2=y_1^2+y_2^2$.
For the spinors and gamma matrices, we use exactly the conventions of \cite{Seibold:2020ywq}, directly replacing the additional $z_3$, $z_4$, $y_3$ and $y_4$ directions of $\mathrm{AdS}_5\times \mathrm{S}^5$ by the four torus fields $x^1,\ldots, x^4$. We then light-cone and $\kappa$ gauge fix as in \cite{Seibold:2020ywq}, ending up with an action depending on the eight bosonic fields $z_1,z_2,y_1,y_2$ and $x^1,\ldots,x^4$, and eight complex Grassmann fields (fermions) labeled $\theta^{13}$, $\theta^{14}$, $\theta^{23}$, $\theta^{24}$, $\eta^{31}$, $\eta^{32}$,  $\eta^{41}$, and $\eta^{42}$ in line with the conventions of \cite{Seibold:2020ywq}. We then expand the action to fourth order in bosons, quadratic order in fermions, forming the starting point for our computation of the tree level S matrices.\footnote{The resulting interaction Lagrangians are too large to practically present them here; they can be found in the Mathematica notebook attached to the arXiv submission for this paper.} We focus on massive excitations, and drop the massless torus bosons $x^1,\ldots,x^4$, and massless fermions $\theta^{13}$, $\theta^{24}$, $\eta^{31}$, and $\eta^{42}$.

\subsection{Common quadratic Lagrangian}

To calculate the S matrix we perform a further change of basis.
We work with the complex bosonic fields
\begin{align}
	Y &= \frac{1}{2 \sqrt{1+\kappa_-^2}} \qty(y_1 + \I y_2)
	\mcomma
	&
	Z &= \frac{1}{2 \sqrt{1+\kappa_-^2}} \qty(z_1 + \I z_2)
	\mcomma
\end{align}
and the complex fermionic fields
\begin{equation}
	\label{field-redefinition-fermions}
	\begin{aligned}
		\zeta_L &= \frac{\massPhasezeta}{\sqrt{2}} \qty(\theta^{14} + \I \theta_{23}^\dagger)
		\mcomma
		\qquad \qquad &
		\chi_L &= -\frac{\massPhasechi}{\sqrt{2}} \qty(\eta^{41} + \I \eta_{32}^\dagger)
		\mcomma
		\\
		\zeta_R &= -\frac{\I}{\sqrt{2}} \qty(\theta^{14} - \I \theta_{23}^\dagger)
		\mcomma
		&
		\chi_R &= -\frac{\I}{\sqrt{2}} \qty(\eta^{41} - \I \eta_{32}^\dagger)
		\mdot
	\end{aligned}
\end{equation}
We introduced the phases $\massPhasezeta$ and $\massPhasechi$ to make all quadratic parts of the Lagrangians of the different deformations take an identical form.
The choices for each of the different deformations are listed in \cref{table-phases-lagrangian}.
The common quadratic part of the Lagrangians is
\begin{equation}
	\label{quadratic-lagrangian}
	\begin{aligned}
		\mathcal{L}_2 = 2 \Big(
			& \abs{\qty(\partial_\tau + \I \kappa_+ \kappa_-) Y}^2
			- \abs{\partial_\sigma Y}^2
			- m^2 \abs{Y}^2
			+ \abs{\qty(\partial_\tau + \I \kappa_+ \kappa_-) Z}^2
			- \abs{\partial_\sigma Z}^2
			- m^2 \abs{Z}^2
		\Big)
		\\
		{} + {} & \I \zeta_L^\dagger \qty(\partial_\tau + \partial_\sigma + \I \kappa_+ \kappa_-) \zeta_L
		+ \I \zeta_R^\dagger \qty(\partial_\tau - \partial_\sigma + \I \kappa_+ \kappa_-) \zeta_R
		- m \zeta_L^\dagger \zeta_R + m \zeta_R^\dagger \zeta_L
		\\
		{} + {} & \I \chi_L^\dagger \qty(\partial_\tau + \partial_\sigma + \I \kappa_+ \kappa_-) \chi_L
		+ \I \chi_R^\dagger \qty(\partial_\tau - \partial_\sigma + \I \kappa_+ \kappa_-) \chi_R
		- m \chi_L^\dagger \chi_R + m \chi_R^\dagger \chi_L
		\mcomma
	\end{aligned}
\end{equation}
where the mass is $m = \sqrt{(1 + \kappa_+^2) (1 + \kappa_-^2)}$.
The field content is two complex bosonic and four complex fermionic fields.
Note that at the level of the quadratic terms the deformation enters as a shift of $\partial_\tau$ by $\pm \I \kappa_+ \kappa_-$ and a deformation of the mass terms.
In contrast, the deformation of the interaction terms takes a more complicated form.

\begin{table}[p]
	\centering
	\begin{tabular}{l@{}cc}
		\toprule
		deformation &
		$\massPhasezeta$ &
		$\massPhasechi$
		\\
		\midrule
		\fermOne{} &
		$1$ &
		$1$
		\\
		\fermTwo{} &
		$1$ &
		$1$
		\\
		\distOne{} &
		$\qty(\frac{\kappa_- - \I}{\kappa_- + \I})^2$ &
		$1$
		\\
		\distTwo{} &
		$\frac{\kappa_- - \I}{\kappa_- + \I} \frac{\kappa_+ - \I}{\kappa_+ + \I}$ &
		$\frac{\kappa_- - \I}{\kappa_- + \I} \frac{\kappa_+ + \I}{\kappa_+ - \I}$
		\\
		\xoxOne{} &
		$-\frac{\kappa_- - \I}{\kappa_- + \I}$ &
		$-\frac{\kappa_- - \I}{\kappa_- + \I}$
		\\
		\xoxTwo{} &
		$1$ &
		$1$
		\\
		\bottomrule
	\end{tabular}
	\caption{%
		Values for the phases $\massPhasezeta$ and $\massPhasechi$ in \cref{field-redefinition-fermions}.
		They are included to ensure that the kinematic Lagrangians of all six cases take the same form \cref{quadratic-lagrangian}.
	}
	\label{table-phases-lagrangian}
\end{table}
\begin{table}[p]
	\centering
	\begin{tabular}{l@{\hspace{-3pt}}ccc@{}c@{\hspace{5pt}}c}
		\toprule
		deformation &
		$\fermionPhase$ &
		$\operatorPhaseYPlusMinus$ &
		$\operatorPhaseZPlusMinus$ &
		$\operatorPhaseZetaPlusMinus$ &
		$\operatorPhaseChiPlusMinus$
		\\
		\midrule
		\fermOne{} &
		$\E^{\I \pi/4} \frac{\sqrt{\kappa_- - \I}}{\sqrt{\kappa_- + \I}}$ &
		$1$ &
		$1$ &
		$1$ &
		$1$
		\\
		\fermTwo{} &
		$\E^{\I \pi/4} \frac{\sqrt{\kappa_- - \I}}{\sqrt{\kappa_- + \I}}$ &
		$1$ &
		$1$ &
		$\begin{aligned}
			\operatorPhase_\pm&(\kappa_+, \kappa_-)
			\\
			& \times \conjugated{\operatorPhase_\pm}(\kappa_-, \kappa_+)
		\end{aligned}$ &
		$\begin{aligned}
			\conjugated{\operatorPhase_\pm}&(\kappa_+, \kappa_-)
			\\
			& \times \operatorPhase_\pm(\kappa_-, \kappa_+)
		\end{aligned}$
		\\
		\distOne{} &
		$\E^{3 \I \pi/4} \frac{\kappa_- - \I}{\kappa_- + \I}$ &
		$1$ &
		$1$ &
		$\operatorPhase_\pm^2(\kappa_+, \kappa_-)$ &
		$1$
		\\
		\distTwo{} &
		$\E^{3 \I \pi/4} \frac{\kappa_- - \I}{\kappa_- + \I}$ &
		$1$ &
		$1$ &
		$\operatorPhase_\pm^2(\kappa_+, \kappa_-)$ &
		$1$
		\\
		\xoxOne{} &
		$\E^{3 \I \pi/4} \frac{\kappa_- - \I}{\kappa_- + \I}$ &
		$1$ &
		$1$ &
		$\operatorPhase_\pm^2(\kappa_+, \kappa_-)$ &
		$1$
		\\
		\xoxTwo{} &
		$\E^{\I \pi/4} \frac{\sqrt{\kappa_- - \I}}{\sqrt{\kappa_- + \I}}$ &
		$\conjugated{\operatorPhase_\pm}(\kappa_-, \kappa_+)$ &
		$\operatorPhase_\pm(\kappa_-, \kappa_+)$ &
		$\operatorPhase_\pm(\kappa_+, \kappa_-)$ &
		$\conjugated{\operatorPhase_\pm}(\kappa_+, \kappa_-)$
		\\
		\bottomrule
	\end{tabular}
	\caption{%
		Values for the phases $\fermionPhase$, $\operatorPhaseYPlusMinus$, $\operatorPhaseZPlusMinus$, $\operatorPhaseZetaPlusMinus$, and $\operatorPhaseChiPlusMinus$ in eq.~\eqref{mode-expansion}.
		We include them to compensate for possible basis differences of the creation and annihilation operators.
		As a result, the T matrices of all six cases take the same form; it is given in \cref{T-matrix}.
		The common factor
		$\operatorPhase_\pm(\kappa_1, \kappa_2) = \frac{\sqrt{p + \I (\kappa_1 \pm \kappa_2 \omega^\pm)}}{\sqrt{p - \I (\kappa_1 \pm \kappa_2 \omega^\pm)}}$
		is the same as the one appearing in this T matrix, see \cref{phase-phi}.
	}
	\label{table-phases-mode-expansion}
\end{table}

Lastly, let us connect the undeformed $\kappa_\pm \to 0$ limit of the quadratic terms to existing results:
Firstly, we reproduce the bosonic and fermionic expressions of \cite{Hoare:2013pma} in the $q \to 0$ limit%
\footnote{Here $q$ controls the mixing of NSNS and RR fluxes, and should not be confused with the $q$ of the quantum deformation discussed in the next section.}
up to a simple normalization of the bosonic fields.
Secondly, when compared to the $\adsFive$ quadratic Lagrangian \cite{Arutyunov:2009ga}, we observe that the bosonic part is a simple truncation, while the fermionic part has inherently different mass terms.

\section{Perturbative S matrix}
\label{sec:Smat_perturbative}
Using the kinetic and interaction part of the gauge-fixed Lagrangian we are now able to calculate the tree-level scattering matrix using Feynman diagram methods.
Firstly, we give the mode expansions of the asymptotic scattering states, and subsequently we calculate the T matrices for all six cases.
Lastly, we show that, similarly to the factorization in the $\adsFive$ case, they can be reduced into smaller building blocks.

From the algebraic perspective -- discussed in the next section -- all six T matrices are expected to be equal, at least up to a unitary change of basis.
However, we want all six T matrices to be exactly identical to simplify comparison with the exact algebraic result.
We achieve this by using the phase freedom of creation and annihilation operators to absorb any possible basis differences with extra phase factors in the mode expansions.

\subsection{On-shell mode expansion}
\label{sec-mode-expansion}
To determine the in- and out-states of the Feynman amplitudes we need to solve the equations of motion for $\mathcal{L}_2$.
They reduce the four complex to four real fermionic degrees of freedom.
Further we express the complex bosonic fields as pairs of real bosonic particles.
In total the particle content of the theory is four real bosons and four real fermions that we will denote by $\{Y_\pm, Z_\pm, \zeta_\pm, \chi_\pm\}$.
They are encoded in the on-shell mode expansions of the original fields, valid for all six cases
\begin{fleqn}
\begin{equation}
	\begin{aligned}
		\label{mode-expansion}
		Y(\tau, \sigma) & =
		\frac{1}{\sqrt{2\pi}} \int \dd{p} \frac{1}{2 \sqrt{\omegaNormalization_p}}
		\qty(
			\E^{\I (p \sigma - \omega_p^+ \tau)} a_{Y_+}(p) \operatorPhaseYPlus
			+ \E^{-\I (p \sigma - \omega_p^- \tau)} a_{Y_-}^\dagger(p) \conjugated{\operatorPhaseYMinus}
		)
		,
		\\
		Z(\tau, \sigma) & =
		\frac{1}{\sqrt{2\pi}} \int \dd{p} \frac{1}{2 \sqrt{\omegaNormalization_p}}
		\qty(
			\E^{\I (p \sigma - \omega_p^+ \tau)} a_{Z_+}(p) \operatorPhaseZPlus
			+ \E^{-\I (p \sigma - \omega_p^- \tau)} a_{Z_-}^\dagger(p) \conjugated{\operatorPhaseZMinus}
		)
		,
		\\
		\zeta_L(\tau, \sigma) & =
		\frac{\fermionPhase}{\sqrt{2\pi}}
		\int \dd{p} \frac{1}{\sqrt{2\omegaNormalization_p}} f_{+p}
		\qty(
			- \E^{\I (p \sigma - \omega_p^+ \tau)} a_{\zeta_+}(p) \operatorPhaseZetaPlus
			+ \E^{-\I (p \sigma - \omega_p^- \tau)} a_{\zeta_-}^\dagger(p) \conjugated{\operatorPhaseZetaMinus}
		)
		,
		\\
		\zeta_R(\tau, \sigma) & =
		\frac{\fermionPhase}{\sqrt{2\pi}}
		\int \dd{p} \frac{1}{\sqrt{2\omegaNormalization_p}} f_{-p}
		\qty(
			- \E^{\I (p \sigma - \omega_p^+ \tau)} a_{\zeta_+}(p) \operatorPhaseZetaPlus
			- \E^{-\I (p \sigma - \omega_p^- \tau)} a_{\zeta_-}^\dagger(p) \conjugated{\operatorPhaseZetaMinus}
		)
		,
		\\
		\chi_L(\tau, \sigma) & =
		\frac{\fermionPhase}{\sqrt{2\pi}}
		\int \dd{p} \frac{1}{\sqrt{2\omegaNormalization_p}} f_{+p}
		\qty(
			\E^{\I (p \sigma - \omega_p^+ \tau)} a_{\chi_+}(p) \operatorPhaseChiPlus
			- \E^{-\I (p \sigma - \omega_p^- \tau)} a_{\chi_-}^\dagger(p) \conjugated{\operatorPhaseChiMinus}
		)
		,
		\\
		\chi_R(\tau, \sigma) & =
		\frac{\fermionPhase}{\sqrt{2\pi}}
		\int \dd{p} \frac{1}{\sqrt{2\omegaNormalization_p}} f_{-p}
		\qty(
			\E^{\I (p \sigma - \omega_p^+ \tau)} a_{\chi_+}(p) \operatorPhaseChiPlus
			+ \E^{-\I (p \sigma - \omega_p^- \tau)} a_{\chi_-}^\dagger(p) \conjugated{\operatorPhaseChiMinus}
		)
		.
	\end{aligned}
\end{equation}
\end{fleqn}
These are complemented by their complex conjugate versions for the respective anti-fields (with $\conjugated{(a_x)} \equiv a_x^\dagger$).
The phases $\fermionPhase$ and $\operatorPhase_{x}$ differ for the six cases and are given in \cref{table-phases-mode-expansion}.
The positive-energy dispersion relation is
\begin{equation}
	\label{perturbative-dispersion-relation}
	\omega_p^\pm = \pm \kappa_+ \kappa_- + \sqrt{p^2 + m^2}
	\mcomma
\end{equation}
with $m^2 = (1 + \kappa_+^2) (1 + \kappa_-^2)$ as before.
The normalization factor $\omegaNormalization_p = \sqrt{p^2 + m^2}$ is chosen such that the worldsheet momentum and energy take their canonical form,
i.e.\ $P_\mup{ws} = \int \dd{p} \sum_x p \, a_x^\dagger a_x$
and $H_\mup{ws} = \int \dd{p} \sum_x \omega_p^x \, a_x^\dagger a_x$,
where $x$ runs over all types of particles $\{Y_\pm, Z_\pm, \zeta_\pm, \chi_\pm\}$.
The wave functions $f_{\pm p} = \sqrt{\pm p + \omegaNormalization_p}$ are fixed by the fermionic equations of motion and by requiring that the quadratic Lagrangian takes the canonical form
\begin{equation}
	\mathcal{L}_2 = \int \dd{p} \sum_x \qty( \I a_x^\dagger(p) \partial_\tau a_x(p) - \omega_p^x a_x^\dagger(p) a_x(p) )
\end{equation}
when inserting the mode expansions with time-dependent creation and annihilation operators into \eqref{quadratic-lagrangian}.

\subsection{Perturbative T matrix}
\label{T-matrix}
Next we calculate the $2 \to 2$ scattering matrix $\dsS$ from the deformed, gauge-fixed and expanded interaction Lagrangian, closely following the notation of \cite{Seibold:2020ywq}.
We expand $\dsS$ in inverse powers of the string tension $\stringtension$ using the tree-level matrix $\dsT$
\begin{equation}
	\dsS = \identity + \frac{\I}{\stringtension} \dsT + \dots
	\mdot
\end{equation}
The procedure of Feynman diagrams provides us then with the tree-level amplitudes.
Details are given in \cref{feynman-diagrammatics} -- here we will only present the results.
But first let us introduce some notation.
The scattering amplitudes depend on two momenta, $p_1$ and $p_2$, with $p_1 > p_2$ by assumption.
The scattering states are
$\ket*{a^\dagger_{x_1}(p_1) a^\dagger_{x_2}(p_2)}
= a^\dagger_{x_1}(p_1) a^\dagger_{x_2}(p_2) \ket{0}$.
To simplify further the states are labeled by their particle content and the first and second particle are taken to depend on $p_1$ and $p_2$ respectively.
For example
\begin{equation}
	\ket{Y_+ \zeta_-}
	\equiv
	\ket{a^\dagger_{Y_+}(p_1) a^\dagger_{\zeta_-}(p_2)}
	, \qquad
	\ket{Z_- \chi_+}
	\equiv
	\ket{a^\dagger_{Z_-}(p_1) a^\dagger_{\chi_+}(p_2)}
	.
\end{equation}
Now all six cases give the same $\dsT$ matrix, as we expect from the algebraic perspective.
Possible unphysical phase discrepancies were prevented by carefully choosing the phases of \cref{table-phases-mode-expansion}.
We give the action of $\dsT$ on the incoming states in the following:

\newcommand{\Aalgebraic}{\mathcal{A}}
\newcommand{\Balgebraic}{\mathcal{B}}
\newcommand{\Galgebraic}{\mathcal{G}}
\newcommand{\Calgebraic}{\mathcal{C}}
\newcommand{\Halgebraic}{\mathcal{H}}
\newcommand{\CbarAlgebraic}{\conjugated{\Calgebraic}}
\newcommand{\HbarAlgebraic}{\conjugated{\Halgebraic}}
\newcommand{\Cinverted}{\CbarAlgebraic}
\newcommand{\Hinverted}{\HbarAlgebraic}
\newcommand{\CbarInverted}{\Calgebraic}
\newcommand{\HbarInverted}{\Halgebraic}

\begingroup
\newcommand{\separationWhitespace}{7pt}
\newcommand{\myintertext}[1]{\intertext{\vskip 5pt \noindent \bfseries \sffamily #1\vspace{10pt}}}
\begin{align*}
\myintertext{Boson-Boson}
	% YY
	\dsT \ket*{Y_\pm Y_\pm} = {}&
		2 (\Aalgebraic + \Balgebraic) \ket*{Y_\pm Y_\pm}
	\\
	\dsT \ket*{Y_\pm Y_\mp} = {}&
		2 \Aalgebraic \ket*{Y_\pm Y_\mp}
		+ \Calgebraic \ket*{\zeta_\pm \zeta_\mp}
		+ \Cinverted \ket*{\chi_\pm \chi_\mp}
	\\
	% ZZ
	\dsT \ket*{Z_\pm Z_\pm} = {}&
		{-2} (\Aalgebraic + \Balgebraic) \ket*{Z_\pm Z_\pm}
	\\
	\dsT \ket*{Z_\pm Z_\mp} = {}&
		{-2} \Aalgebraic \ket*{Z_\pm Z_\mp}
		- \CbarInverted \ket*{\zeta_\pm \zeta_\mp}
		- \CbarAlgebraic \ket*{\chi_\pm \chi_\mp}
	\\[\separationWhitespace]
	% YZ
	\dsT \ket*{Y_\pm Z_\pm} = {}&
		2 \Galgebraic \ket*{Y_\pm Z_\pm}
		- \Halgebraic \ket*{\zeta_\pm \chi_\pm}
		+ \Hinverted \ket*{\chi_\pm \zeta_\pm}
	\\
	\dsT \ket*{Y_\pm Z_\mp} = {}&
		2 \Galgebraic \ket*{Y_\pm Z_\mp}
	\\
	% ZY
	\dsT \ket*{Z_\pm Y_\pm} = {}&
		{-2} \Galgebraic \ket*{Z_\pm Y_\pm}
		- \HbarInverted \ket*{\zeta_\pm \chi_\pm}
		+ \HbarAlgebraic \ket*{\chi_\pm \zeta_\pm}
	\\
	\dsT \ket*{Z_\pm Y_\mp} = {}&
		{-2} \Galgebraic \ket*{Z_\pm Y_\mp}
\myintertext{Fermion-Fermion}
	% χχ
	\dsT \ket*{\zeta_\pm \zeta_\mp} = {}&
		\CbarAlgebraic \ket*{Y_\pm Y_\mp}
		- \Cinverted \ket*{Z_\pm Z_\mp}
	\\
	% ζζ
	\dsT \ket*{\chi_\pm \chi_\mp} = {}&
		\CbarInverted \ket*{Y_\pm Y_\mp}
		- \Calgebraic \ket*{Z_\pm Z_\mp}
	\\
	% χζ
	\dsT \ket*{\zeta_\pm \chi_\pm} = {}&
		{- \HbarAlgebraic} \ket*{Y_\pm Z_\pm}
		- \Hinverted \ket*{Z_\pm Y_\pm}
	\\
	% ζχ
	\dsT \ket*{\chi_\pm \zeta_\pm} = {}&
		\HbarInverted \ket*{Y_\pm Z_\pm}
		+ \Halgebraic \ket*{Z_\pm Y_\pm}
\myintertext{Boson-Fermion}
	% Yχ
	\dsT \ket*{Y_\pm \zeta_\pm} = {}&
		(\Aalgebraic + \Balgebraic + \Galgebraic) \ket*{Y_\pm \zeta_\pm}
		+ \Halgebraic \ket*{\zeta_\pm Y_\pm}
	\\
	\dsT \ket*{Y_\pm \zeta_\mp} = {}&
		(\Aalgebraic + \Galgebraic) \ket*{Y_\pm \zeta_\mp}
		+ \Cinverted \ket*{\chi_\pm Z_\mp}
	\\
	% Yζ
	\dsT \ket*{Y_\pm \chi_\pm} = {}&
		(\Aalgebraic + \Balgebraic + \Galgebraic) \ket*{Y_\pm \chi_\pm}
		+ \Hinverted \ket*{\chi_\pm Y_\pm}
	\\
	\dsT \ket*{Y_\pm \chi_\mp} = {}&
		(\Aalgebraic + \Galgebraic) \ket*{Y_\pm \chi_\mp}
		- \Calgebraic \ket*{\zeta_\pm Z_\mp}
	\\[\separationWhitespace]
	% Zχ
	\dsT \ket*{Z_\pm \zeta_\pm} = {}&
		(-\Aalgebraic - \Balgebraic - \Galgebraic) \ket*{Z_\pm \zeta_\pm}
		- \HbarInverted \ket*{\zeta_\pm Z_\pm}
	\\
	\dsT \ket*{Z_\pm \zeta_\mp} = {}&
		(-\Aalgebraic - \Galgebraic) \ket*{Z_\pm \zeta_\mp}
		+ \CbarAlgebraic \ket*{\chi_\pm Y_\mp}
	\\
	% Zζ
	\dsT \ket*{Z_\pm \chi_\pm} = {}&
		(-\Aalgebraic - \Balgebraic - \Galgebraic) \ket*{Z_\pm \chi_\pm}
		- \HbarAlgebraic \ket*{\chi_\pm Z_\pm}
	\\
	\dsT \ket*{Z_\pm \chi_\mp} = {}&
		(-\Aalgebraic - \Galgebraic) \ket*{Z_\pm \chi_\mp}
		- \CbarInverted \ket*{\zeta_\pm Y_\mp}
	\\[\separationWhitespace]
	% χY
	\dsT \ket*{\zeta_\pm Y_\pm} = {}&
		(\Aalgebraic + \Balgebraic - \Galgebraic) \ket*{\zeta_\pm Y_\pm}
		+ \HbarAlgebraic \ket*{Y_\pm \zeta_\pm}
	\\
	\dsT \ket*{\zeta_\pm Y_\mp} = {}&
		(\Aalgebraic - \Galgebraic) \ket*{\zeta_\pm Y_\mp}
		- \Cinverted \ket*{Z_\pm \chi_\mp}
	\\
	% χZ
	\dsT \ket*{\zeta_\pm Z_\pm} = {}&
		(-\Aalgebraic - \Balgebraic + \Galgebraic) \ket*{\zeta_\pm Z_\pm}
		- \Hinverted \ket*{Z_\pm \zeta_\pm}
	\\
	\dsT \ket*{\zeta_\pm Z_\mp} = {}&
		(-\Aalgebraic + \Galgebraic) \ket*{\zeta_\pm Z_\mp}
		- \CbarAlgebraic \ket*{Y_\pm \chi_\mp}
	\\[\separationWhitespace]
	% ζY
	\dsT \ket*{\chi_\pm Y_\pm} = {}&
		(\Aalgebraic + \Balgebraic - \Galgebraic) \ket*{\chi_\pm Y_\pm}
		+ \HbarInverted \ket*{Y_\pm \chi_\pm}
	\\
	\dsT \ket*{\chi_\pm Y_\mp} = {}&
		(\Aalgebraic - \Galgebraic) \ket*{\chi_\pm Y_\mp}
		+ \Calgebraic \ket*{Z_\pm \zeta_\mp}
	\\
	% ζZ
	\dsT \ket*{\chi_\pm Z_\pm} = {}&
		(-\Aalgebraic - \Balgebraic + \Galgebraic) \ket*{\chi_\pm Z_\pm}
		- \Halgebraic \ket*{Z_\pm \chi_\pm}
	\\
	\dsT \ket*{\chi_\pm Z_\mp} = {}&
		(-\Aalgebraic + \Galgebraic) \ket*{\chi_\pm Z_\mp}
		+ \CbarInverted \ket*{Y_\pm \zeta_\mp}
\end{align*}
\endgroup

\noindent
We only work to second order in fermions and therefore our result does not contain four-fermion amplitudes.
The coefficients all carry two implicit indices -- omitted for brevity -- that match the $\pm$ indices of the state they stand next to.
For example the last line contains coefficients $\Aalgebraic_{\pm \mp}$, $\Galgebraic_{\pm \mp}$, and $\Calgebraic_{\pm \mp}$.
The coefficients are
\begin{fleqn}
\begin{equation}
	\label{T-matrix-coefficients}
	\begin{aligned}
		\Aalgebraic_{\mu_1 \mu_2} &=
		\frac{1}{4 D} \big( (p_1 - p_2)^2 + (\kappa_+^2 + \kappa_-^2)(\omega_1 - \omega_2)^2 + 2 \kappa_+ \kappa_- (\mu_1 \mu_2 \omega_1 \omega_2 + 1)(\mu_1 \omega_1 + \mathrlap{\mu_2 \omega_2) \big)}
		\\
		\Balgebraic_{\mu_1 \mu_2} &=
			\frac{1}{D} \qty(p_1 p_2 + (\kappa_+^2 + \kappa_-^2) \omega_1 \omega_2)
		\\
		\Galgebraic_{\mu_1 \mu_2} &=
			\frac{1}{4 D} \qty( -(1 + \kappa_+^2 + \kappa_-^2)(p_1^2 - p_2^2) - 2 \kappa_+ \kappa_- (p_1^2 \mu_2 \omega_2 - p_2^2 \mu_1 \omega_1) )
		\\
		\Calgebraic_{\mu_1 \mu_2} &=
			- \conjugated{\varphi_1} \conjugated{\varphi_2} \frac{m}{p_1 + p_2} \sqrt{\omega_1^2 - 1} \sqrt{\omega_2^2 - 1} \sinh( \frac{1}{2} \qty( \arsinh \frac{p_1}{m} + \arsinh \frac{p_2}{m} ) )
		\\
		\Halgebraic_{\mu_1 \mu_2} &=
			\phantom{+} \conjugated{\varphi_1} \varphi_2 \frac{m}{p_1 - p_2} \sqrt{\omega_1^2 - 1} \sqrt{\omega_2^2 - 1} \cosh( \frac{1}{2} \qty( \arsinh \frac{p_1}{m} + \arsinh \frac{p_2}{m} ) )
	\end{aligned}
\end{equation}
\end{fleqn}
where $\mu_{1,2} = \pm$ and $\omega_i \equiv \omega_{p_i}^{\mu_i}$, see \cref{perturbative-dispersion-relation}.
In addition $D = p_1 \omega_2 - p_2 \omega_1$ and the phase
\begin{equation}
	\label{phase-phi}
	\varphi_j = \frac{\sqrt{p_j + \I (\kappa_+ + \mu_j \kappa_- \omega_j)}}
	         {\sqrt{p_j - \I (\kappa_+ + \mu_j \kappa_- \omega_j)}}
	\mdot
\end{equation}
The operator $\dsT$ satisfies the classical Yang-Baxter equation%
\footnote{
	\label{definition-graded-embeddings-T}%
	The $\dsT_{ij}$ are the graded embeddings (using the graded permutation operator $P^\mup{g}_{ij}$, defined, for example, in eq.\ (3.8) of \cite{Arutyunov:2009ga}) of $\dsT$ into the product of three spaces.
	Explicitly these are
	\begin{equation*}
		\dsT_{12} = \dsT \otimes \identity
		\mcomma \qquad
		\dsT_{13} = P^\mup{g}_{23} \dsT_{12} P^\mup{g}_{23}
		\mcomma \qquad
		\dsT_{23} = P^\mup{g}_{12} P^\mup{g}_{13} \dsT_{12} P^\mup{g}_{13} P^\mup{g}_{12} = \identity \otimes \dsT
		\mdot
	\end{equation*}
}
\begin{equation}
	\commutator{\dsT_{23}}{\dsT_{13}}
	+ \commutator{\dsT_{23}}{\dsT_{12}}
	+ \commutator{\dsT_{13}}{\dsT_{12}}
	= 0
\end{equation}
for all terms that our restriction to processes with at most two fermions allowed us to check.
Furthermore, the bosonic amplitudes match the two-parameter limit of the amplitudes computed in \cite{Bocconcello:2020qkt}.

We want to comment on the phase factors $\varphi_j$ appearing in the final amplitudes.
These factors are directly influenced by the extra phases in the mode expansions, see \cref{table-phases-mode-expansion}.
We fixed these phases such that the perturbative result matches the algebraic result.
However, as we will see in \cref{sec:Smat_exact}, the algebraic result cannot fix the phases completely; there is still the freedom to choose the phase $\alpha$, see \cref{definition-gamma}.
In principle we could use this freedom and choose the extra phases such that the phase factors $\varphi_j$ would disappear, as is common practice in previous works on deformed S matrices.
However, this would introduce either a discontinuity in the phase of the amplitudes for non-zero $\kappa_\pm$ or a sign (branch choice) mismatch with the undeformed result.
To avoid these complications, we refrain from modifying the extra phases and accept the final amplitudes to be complex.

\subsection{Factorization}
\label{T-matrix-factorization}
Similarly to the $\adsFive$ case, it is possible to write the $\dsT$ matrix in a factorized form.
For this we write the single particle states as product states of two of the $\mathfrak{psu}_q(1|1)^2_\mup{c.e.}$ representations considered in the next section.
We further index the bases $\{ \phi_\pm, \psi_\pm \}$ of the two representations (see \cref{fundamental-q-def-S}) from $1$ to $4$ for the first and $\dot{1}$ to $\dot{4}$ for the second copy.
We get
\begin{equation}
	\label{particle-index-mapping}
	\begin{aligned}
		Y_+ & = \phi_+ \otimes \phi_+ \leftrightarrow 1 \dot{1}
		\mcomma
		\qquad \qquad &
		Z_+ & = \psi_+ \otimes \psi_+ \leftrightarrow 3 \dot{3}
		\mcomma
		\\
		Y_- & = \phi_- \otimes \phi_- \leftrightarrow 2 \dot{2}
		\mcomma
		&
		Z_- & = \psi_- \otimes \psi_- \leftrightarrow 4 \dot{4}
		\mcomma
		\\
		\zeta_+ & = \phi_+ \otimes \psi_+ \leftrightarrow 1 \dot{3}
		\mcomma
		&
		\chi_+ & = \psi_+ \otimes \phi_+ \leftrightarrow 3 \dot{1}
		\mcomma
		\\
		\zeta_- & = \phi_- \otimes \psi_- \leftrightarrow 2 \dot{4}
		\mcomma
		&
		\chi_- & = \psi_- \otimes \phi_- \leftrightarrow 4 \dot{2}
		\mdot
	\end{aligned}
\end{equation}
The indices allow us to express the matrix elements of $\dsT$ as
\begin{equation}
	\label{factorized-T-matrix-indices}
	\begin{aligned}
		\dsT^{P \dot{P} Q \dot{Q}}_{M \dot{M} N \dot{N}}
		=
		(-1)^{\epsilon_{\dot{M}} (\epsilon_N + \epsilon_Q)}
		&
		\scT^{PQ}_{MN}(-\kappa_+, -\kappa_-) \delta^{\dot{P}}_{\dot{M}} \delta^{\dot{Q}}_{\dot{N}}
		\\
		& {} +
		(-1)^{\epsilon_Q (\epsilon_{\dot{M}} + \epsilon_{\dot{P}})}
		\delta^P_M \delta^Q_N \scT^{\dot{P}\dot{Q}}_{\dot{M}\dot{N}}(\kappa_+, \kappa_-)
	\end{aligned}
\end{equation}
up to the four-fermion amplitudes that we did not compute.
The $\scT^{PQ}_{MN}$ are the matrix elements of the expansion of the exact $\mathfrak{psu}_q(1|1)^2_\mup{c.e.}$ S matrix that will be derived in the next section.
$\epsilon_M$ is $0$ if $\scriptstyle M$ is a bosonic index (1 or 2) and $1$ if it is a fermionic index (3 or 4).
Note that the pairs of undotted and dotted indices only take the values
$1 \dot{1}$, $2 \dot{2}$, $3 \dot{3}$, $4 \dot{4}$, $1 \dot{3}$, $2 \dot{4}$, $3 \dot{1}$, and $4 \dot{2}$.
\newcommand{\truncate}[1]{\qty[#1]_\mup{trunc}}
If we denote the operation of truncating a matrix to the subspace spanned by these indices with $\truncate{\:\cdot\:}$, then we can express $\dsT$ in the basis-independent form
\begin{equation}
	\dsT = \truncate{\scT(-\kappa_+, -\kappa_-) \otimes \identity + \identity \otimes \scT(\kappa_+, \kappa_-)}
	\mdot
\end{equation}
The factor $\scT(\kappa_+, \kappa_-)$ is specified by%
\footnote{%
	Latin indices take values $1$ and $2$, Greek indices take values $3$ and $4$, and unspecified entries are zero.
	The coefficients are defined above in \cref{T-matrix-coefficients}.
	Again we suppressed their two indices.
	The first index of each coefficient is $+$ if $a = 1$ or $\alpha = 3$ and $-$ if $a = 2$ or $\alpha = 4$, cf.~\cref{particle-index-mapping}.
	The second index depends in the same manner on $b$ or $\beta$.
}
\begin{equation}
	\label{definition-scT}
	\begin{aligned}
		& \scT_{ab}^{cd} = \Aalgebraic \delta_a^c \delta_b^d + \Balgebraic \delta_{ab} \delta_a^d \delta_b^c
		\mcomma \\
		& \scT_{\alpha \beta}^{\gamma \delta} = -\Aalgebraic \delta_\alpha^\gamma \delta_\beta^\delta
			-\Balgebraic \delta_{\alpha \beta} \delta_\alpha^\delta \delta_\beta^\gamma
		\mcomma \\
		& \begin{alignedat}{2}
			\scT_{a \beta}^{c \delta} & = \Galgebraic \delta_a^c \delta_\beta^\delta
			\mcomma \qquad \quad &
			\scT_{\alpha b}^{\gamma d} & = -\Galgebraic \delta_\alpha^\gamma \delta_b^d
			\mcomma \\
			\scT_{12}^{34} & = \scT_{21}^{43} = \Calgebraic
			\mcomma &
			\scT_{34}^{12} & = \scT_{43}^{21} = \CbarAlgebraic
			\mcomma \\
			\scT_{13}^{31} & = \scT_{24}^{42} = \Halgebraic
			\mcomma &
			\scT_{31}^{13} & = \scT_{42}^{24} = \HbarAlgebraic
			\mdot
		\end{alignedat}
	\end{aligned}
\end{equation}

The sign inversion in the first factor of \cref{factorized-T-matrix-indices} effectively behaves like a complex conjugation, $\scT(-\kappa_+, -\kappa_-) = \conjugated{\scT(\kappa_+, \kappa_-)}$.
This only affects the complex coefficients $\Calgebraic$ and $\Halgebraic$, whose phases $\varphi_j$ get conjugated -- everything else is invariant.

In the undeformed limit $\kappa_\pm \to 0$ the $\dsT$ matrix matches the $q = 0$ case of \cite{Hoare:2013pma}.
In the one-parameter limit $\kappa_- \to 0$ and when identifying $\kappa_+ \equiv \kappa$ the matrix $\dsT$ matches the truncation of the tree-level $\dsT$ matrix of $\eta$-deformed $\adsFive$ of \cite{Seibold:2020ywq} up to phases of the terms containing $\Calgebraic$ and $\Halgebraic$, even though the structure of the $\scT$'s differ.

\section{The exact q-deformed S matrix}
\label{sec:Smat_exact}
In the previous section we obtained the tree-level contribution to the massive two-body scattering matrix using perturbation theory. In fact, the all-loop scattering matrix can in some cases be bootstrapped (at least partially, up to the dressing factors) using the symmetries of the light-cone gauge fixed theory. This is in particular true for the $\adsThreeT$ superstring~\cite{Borsato:2013qpa}. The two-parameter integrable deformations of the $\adsThreeT$ superstring that we consider are expected to correspond to quantum deformations, for which the symmetry algebra of the original theory is promoted to a quantum group~\cite{Delduc:2014kha,Hoare:2014oua}. It turns out that these quantum-deformed symmetries are powerful enough to also bootstrap the scattering matrix in the deformed case~\cite{Beisert:2008tw,Seibold:2020ywq} and a proposal for the doubly $q$-deformed S matrix has been obtained in \cite{Hoare:2014oua}.

In this section we review the symmetries of the various two-parameter integrable deformations of the $\adsThreeT$ superstring, as well as of their light-cone gauge fixed theories, and recall the expression of the expected exact S matrix. We then expand the latter in inverse powers of the string tension and match the tree-level contribution with the perturbative results obtained in the previous section.

\subsection{Symmetries of the deformed theories}
The symmetry algebra of the $\adsThree$ superstring is given by two copies of the $\mathfrak{psu}(1,1|2)$ superalgebra. It contains the symmetry algebra of $\AdS_3$ ($\mathfrak{so}(2,2) \cong \mathfrak{su}(1,1) \oplus \mathfrak{su}(1,1)$) and of the three-sphere ($\mathfrak{so}(4) \cong \mathfrak{su}(2) \oplus \mathfrak{su}(2)$) as (bosonic) subalgebras. Upon light-cone gauge fixing and taking the decompactification limit to have well-defined asymptotic scattering states, this symmetry algebra becomes
\begin{equation}
\label{eq:sym_alg_AdS3}
\left[\mathfrak{psu}(1|1) \oplus \mathfrak{psu}(1|1) \right]^{\oplus 2}_{c.e.}~.
\end{equation}
Each of the two $\mathfrak{psu}(1,1|2)$ copies is broken down to two $\mathfrak{psu}(1|1)$ factors and the symmetry algebra gets enhanced by four central charges.
The exact scattering matrix can then be bootstrapped using these symmetries. More precisely, we impose
\begin{equation}
\Delta_\text{op}(X) \,\dsS = \dsS\, \Delta(X)~,
\end{equation}
where $X$ is an element of the symmetry algebra \eqref{eq:sym_alg_AdS3}, $\Delta$ denotes the coproduct that turns this symmetry algebra into a Hopf algebra, and $\Delta_\text{op}$ is the opposite coproduct.\footnote{In the presence of central elements the coproduct is deformed by introducing the braiding, so as to obtain a non-trivial S matrix.} As can be readily seen, this equation leaves the freedom to choose an overall prefactor. In fact, due to the structure of \eqref{eq:sym_alg_AdS3} there will be four possible overall prefactors and these are the four dressing factors. They are constrained by requiring the S matrix to be braiding unitary, matrix unitary and crossing symmetric,
\begin{equation}
\label{eq:Smat_phys_cond}
\dsS_{12} \dsS_{21} =1~,  \qquad \dsS^\dagger \dsS = 1~, \qquad (\mathcal{C}^{-1} \otimes 1) \dsS^{\supertranspose \otimes 1}_{\bar{1}2} (\mathcal C \otimes 1) \dsS_{12}=1~,
\end{equation}
where $\mathcal C$ is the charge conjugation matrix, $^\supertranspose$ denotes the super-transposition (see for instance \cite{Beisert:2008tw} or \cref{foot:ST}) and $\bar{1}$ denotes the antipode representation, encoding the particle to antiparticle transformation.

For the various two-parameter integrable deformations of the $\adsThree$ superstring, the symmetry algebra is conjectured to correspond to a quantum group~\cite{Hoare:2014oua}
\begin{equation}
\mathfrak{psu}_{q_L}(1,1|2) \oplus \mathfrak{psu}_{q_R}(1,1|2)~,
\end{equation}
with two different parameters $q_L$ and $q_R$ in the two copies, allowing for asymmetric deformations. The case $q_L=q_R$ corresponds to the one-parameter $\eta$-deformations, and the limit $q_L = q_R \rightarrow 1$ sends them all to the undeformed theory. Quantum groups are defined through the choice of a Cartan-Weyl basis. For superalgebras, not all Cartan-Weyl bases are equivalent, the inequivalent choices are characterized by different Dynkin diagrams. Quantum groups associated to different Dynkin diagrams will have different properties. In particular, the $\mathfrak{psu}(1,1|2)$ algebra admits three different Dynkin diagrams
\begin{equation}
\Circle - \otimes - \Circle \qquad \otimes - \Circle - \otimes \qquad \otimes - \otimes - \otimes~,
\end{equation}
with $\Circle$ and $\otimes$ denoting bosonic and fermionic simple roots respectively.
This is why we have different deformed models to start with. Also the symmetry algebra of the light-cone gauge fixed theory is assumed to correspond to a quantum group. An analysis similar to the one performed in the $\adsFive$ case indicates that the two copies are deformed in an opposite way, leading to the light-cone symmetry algebra
\begin{equation}
\big[\mathfrak{psu}_{q_L^{-1}}(1|1) \oplus \mathfrak{psu}(1|1)_{q_R^{-1}} \big]_{c.e.}  \oplus \big[\mathfrak{psu}_{q_L}(1|1) \oplus \mathfrak{psu}(1|1)_{q_R} \big]_{c.e.}~.
\end{equation}
The fact that the symmetry algebra factorizes into two copies linked by the central elements (which are the same for both copies) indicates that also the S matrix will factorize,
\begin{equation}
\mathbb{S}= \mathcal{S}_{q_L^{-1}, q_R^{-1}} \, \otimes \, \mathcal{S}_{q_L, q_R}~,
\end{equation}
where the factorized $q$-deformed S matrix $\mathcal{S}_{q_L, q_R}$ is invariant under
\begin{equation}
\label{eq:sym_alg_fac}
\left[\mathfrak{psu}_{q_L}(1|1) \oplus \mathfrak{psu}(1|1)_{q_R} \right]_{c.e.}~.
\end{equation}
As noted previously, quantum groups based on different Dynkin diagrams are not equivalent. Their coproducts are related by a twist and hence also the S matrices will be related by a twist. This is precisely what happened in the case of the $\eta$-deformed $\adsFive$ superstring~\cite{Seibold:2020ywq}, whose light-cone gauge fixed symmetry algebra is given by two copies of centrally extended $q$-deformed $\mathfrak{psu}(2|2)$. In contrast, $\mathfrak{psu}(1|1)$ is a rank one superalgebra with a unique Dynkin diagram, formed by a single fermionic simple root $\otimes$. We thus expect all the deformations to have the same factorized scattering matrix, up to one-particle change of basis and possibly different dressing factors.

\begin{figure}
\begin{subfigure}[b]{0.5 \textwidth}
\begin{equation*}
\addtolength{\arraycolsep}{-1pt}
\left(
\begin{array}{cccc}
C \cellcolor{color3}  &  + & + \cellcolor{color3} & + \\
- &  C \cellcolor{color4} & -  & + \cellcolor{color4} \\
- \cellcolor{color3} & + & C \cellcolor{color3} & + \\
-  & - \cellcolor{color4} & - & C\cellcolor{color4} \\
\end{array}
\right)
\oplus
\left(
\begin{array}{cccc}
C \cellcolor{color3}  &  - & - \cellcolor{color3} & - \\
+ &  C \cellcolor{color4} & +  & - \cellcolor{color4} \\
+ \cellcolor{color3} & - & C \cellcolor{color3} & - \\
+  & + \cellcolor{color4} & + & C\cellcolor{color4} \\
\end{array}
\right)
\end{equation*}
\caption{\fermOne{}}
\end{subfigure}
\hfill
\begin{subfigure}[b]{0.5 \textwidth}
\begin{equation*}
\addtolength{\arraycolsep}{-1pt}
\left(
\begin{array}{cccc}
C \cellcolor{color3}  &  + & + \cellcolor{color3} & + \\
- &  C \cellcolor{color4} & -  & + \cellcolor{color4} \\
- \cellcolor{color3} & + & C \cellcolor{color3} & + \\
-  & - \cellcolor{color4} & - & C\cellcolor{color4} \\
\end{array}
\right)
\oplus
\left(
\begin{array}{cccc}
C \cellcolor{color3}  &  - & + \cellcolor{color3} & - \\
+ &  C \cellcolor{color4} & +  & + \cellcolor{color4} \\
- \cellcolor{color3} & - & C \cellcolor{color3} & - \\
+  & - \cellcolor{color4} & + & C\cellcolor{color4} \\
\end{array}
\right)
\end{equation*}
\caption{\fermTwo{}}
\end{subfigure}

\bigskip
\begin{subfigure}[b]{0.5 \textwidth}
\begin{equation*}
\addtolength{\arraycolsep}{-1pt}
\left(
\begin{array}{cccc}
C \cellcolor{color3}  &  + & + \cellcolor{color3} & + \\
- &  C \cellcolor{color4} & +  & + \cellcolor{color4} \\
- \cellcolor{color3} & - & C \cellcolor{color3} & + \\
-  & - \cellcolor{color4} & - & C\cellcolor{color4}
\end{array}
\right)
\oplus
\left(
\begin{array}{cccc}
C \cellcolor{color3}  &  - & - \cellcolor{color3} & - \\
+ &  C \cellcolor{color4} & -  & - \cellcolor{color4} \\
+ \cellcolor{color3} & + & C \cellcolor{color3} & - \\
+  & + \cellcolor{color4} & + & C\cellcolor{color4}
\end{array}
\right)
\end{equation*}
\caption{\distOne{}}
\end{subfigure}
\hfill
\begin{subfigure}[b]{0.5 \textwidth}
\begin{equation*}
\addtolength{\arraycolsep}{-1pt}
\left(
\begin{array}{cccc}
C \cellcolor{color3}  &  + & + \cellcolor{color3} & + \\
- &  C \cellcolor{color4} & +  & + \cellcolor{color4} \\
- \cellcolor{color3} & - & C \cellcolor{color3} & + \\
-  & - \cellcolor{color4} & - & C\cellcolor{color4}
\end{array}
\right)
\oplus
\left(
\begin{array}{cccc}
C \cellcolor{color3}  &  - & + \cellcolor{color3} & + \\
+ &  C \cellcolor{color4} & +  & + \cellcolor{color4} \\
- \cellcolor{color3} & - & C \cellcolor{color3} & - \\
-  & - \cellcolor{color4} & + & C\cellcolor{color4}
\end{array}
\right)
\end{equation*}
\caption{\distTwo{}}
\end{subfigure}

\bigskip
\begin{subfigure}[b]{0.5 \textwidth}
\begin{equation*}
\addtolength{\arraycolsep}{-1pt}
\left(
\begin{array}{cccc}
C \cellcolor{color3}  &  + & + \cellcolor{color3} & + \\
- &  C \cellcolor{color4} & -  & - \cellcolor{color4} \\
- \cellcolor{color3} & + & C \cellcolor{color3} & + \\
-  & + \cellcolor{color4} & - & C\cellcolor{color4}
\end{array}
\right)
\oplus
\left(
\begin{array}{cccc}
C \cellcolor{color3}  &  - & - \cellcolor{color3} & - \\
+ &  C \cellcolor{color4} & +  & + \cellcolor{color4} \\
+ \cellcolor{color3} & - & C \cellcolor{color3} & - \\
+  & - \cellcolor{color4} & + & C\cellcolor{color4}
\end{array}
\right)
\end{equation*}
\caption{\xoxOne{}}
\end{subfigure}
\hfill
\begin{subfigure}[b]{0.5 \textwidth}
\begin{equation*}
\addtolength{\arraycolsep}{-1pt}
\left(
\begin{array}{cccc}
C \cellcolor{color3}  &  + & + \cellcolor{color3} & + \\
- &  C \cellcolor{color4} & -  & - \cellcolor{color4} \\
- \cellcolor{color3} & + & C \cellcolor{color3} & + \\
-  & + \cellcolor{color4} & - & C\cellcolor{color4}
\end{array}
\right)
\oplus
\left(
\begin{array}{cccc}
C \cellcolor{color3}  &  - & + \cellcolor{color3} & - \\
+ &  C \cellcolor{color4} & +  & - \cellcolor{color4} \\
- \cellcolor{color3} & - & C \cellcolor{color3} & - \\
+  & + \cellcolor{color4} & + & C\cellcolor{color4}
\end{array}
\right)
\end{equation*}
\caption{\xoxTwo{}}
\end{subfigure}
\caption{Structure of an element in $\mathfrak{psu}(1,1|2)^{\oplus 2}$. The blue and red elements each generate a $\mathfrak{psu}(1|1)^{\oplus 2}$ subalgebra. After deformation the symmetry algebra is promoted to a quantum group associated to the Cartan-Weyl basis $(C,+,-)$ where $C$ denotes the Cartan elements, $+$ denotes the positive roots and $-$ denotes the negative roots. The choice of Cartan-Weyl basis differs for the different deformations. When going from \fermOne{} to \fermTwo{} (or \distOne{} to \distTwo{}, \xoxOne{} to \xoxTwo{}), positive and negative roots are exchanged in the second copy. The two deformations are thus related by an inversion $q_R \rightarrow q_R^{-1}$.}
\label{fig:q-def}
\end{figure}

Even though there is a unique Dynkin diagram one can still choose different Cartan-Weyl bases in the two $\mathfrak{psu}(1|1)$ copies. These bases are related by Weyl transformations and one expects the resulting S matrices to be related by a one-particle change of basis. This is what happens when going from the \fermOne{} to the \fermTwo{} background for instance. By looking at the action of the Drinfel'd-Jimbo operator $R$ in the Lagrangian of the two-parameter deformations one sees that between \fermOne{} and \fermTwo{} the notion of positive and negative root is exchanged in the second copy, see Figure~\ref{fig:q-def}. One thus expects the two factorized S matrices to be related by $q_R \rightarrow q_R^{-1}$. We will see later that this is indeed the case and it can be reabsorbed into a one-particle change of basis. The same holds for the distinction between the two distinguished and \xox{} backgrounds.

Finally let us note that quantum integrability imposes strong constraints on the S matrix $\dsS$. The set of incoming momenta should be the same as the set of outgoing momenta (in particular there cannot be any particle production) and an $n$-body scattering event should be decomposable into a sequence of two-body scattering events. This entails that it is sufficient to know the two-body S matrix, and consistency requires the latter to satisfy the quantum Yang-Baxter equation.

\subsection{Fundamental q-deformed S matrix}
\label{fundamental-q-def-S}
The fundamental S matrix compatible with the symmetry algebra \eqref{eq:sym_alg_fac} has been constructed and analyzed in \cite{Hoare:2014oua}. Here we recall its expression and its main properties. Let us consider two different two-dimensional representations $\{\ket{\phi_+}, \ket{\psi_+}\}$ and $\{\ket{\phi_-}, \ket{\psi_-}\}$ of \eqref{eq:sym_alg_fac} with $q_L \in \mathbb{R}_{>0, \neq 1}$ and $q_R \in \mathbb{R}_{>0, \neq 1}$. Upon removing the central elements, $\{\ket{\phi_+}, \ket{\psi_+}\}$ transforms in the fundamental representation of one $\mathfrak{psu}(1|1)$ copy, while $\{\ket{\phi_-}, \ket{\psi_-}\}$ forms the fundamental representation of the other $\mathfrak{psu}(1|1)$ copy.%
\footnote{This is why in the literature these are sometimes also denoted by $\{\ket{\phi_L}, \ket{\psi_L}\}$ and $\{\ket{\phi_R}, \ket{\psi_R}\}$, but the notation with $\pm$ will allow us to simplify expressions.}
The central elements will however non-trivially mix the two representations. The states $\ket{\phi_\pm}$ are bosonic, while the states  $\ket{\psi_\pm}$ are fermionic. The S matrix acts as%
\footnote{A brief comment about notation. In harmony with the previous sections we use $\dsS$ for the full S matrix, $\scS$ for the factorized S matrix (including the dressing factors) and $\scR$ for the factorized S matrix with a specific choice of dressing factors. In particular, $\scR$ does not need to satisfy braiding unitarity, matrix unitarity and crossing symmetry. Schematically we have $\scS = R \scR$ with $R$ the dressing factors.}
\begin{align}
\label{eq:Smat}
\scR \ket{\phi_\pm \phi_\pm} &= A_{\pm \pm} \ket{\phi_\pm \phi_\pm} &\qquad  \scR \ket{\psi_\pm \psi_\pm} &= F_{\pm \pm} \ket{\psi_\pm \psi_\pm} \\
\nonumber
\scR \ket{\phi_\pm \psi_\pm} &= B_{\pm \pm} \ket{\phi_\pm \psi_\pm}+C_{\pm \pm} \ket{\psi_\pm \phi_\pm}  &\qquad  \scR \ket{\psi_\pm \phi_\pm} &= D_{\pm \pm} \ket{\psi_\pm \phi_\pm}+ E_{\pm \pm}\ket{\phi_\pm \psi_\pm} \\
\nonumber
\scR \ket{\phi_\pm \phi_\mp} &= A_{\pm \mp} \ket{\phi_\pm \phi_\mp} + B_{\pm \mp} \ket{\psi_\pm \psi_\mp} &\qquad  \scR \ket{\psi_\pm \psi_\mp} &= E_{\pm \mp} \ket{\psi_\pm \psi_\mp} + F_{\pm \mp} \ket{\phi_\pm \phi_\mp} \\
\nonumber
\scR \ket{\phi_\pm \psi_\mp} &= C_{\pm \mp} \ket{\phi_\pm \psi_\mp}  &\qquad  \scR \ket{\psi_\pm \phi_\mp} &= D_{\pm \mp} \ket{\psi_\pm \phi_\mp}
\end{align}
with coefficients given by
\begin{fleqn}
\begin{equation}
\begin{aligned}
\label{eq:Smat_coeffs}
A_{\pm \pm} &= \frac{U_1 V_1 W_1}{U_2 V_2 W_2} \frac{x_1^- - x_2^+}{x_1^+ - x_2^-}, &\quad F_{\pm \pm} &= 1 \mcomma \\
B_{\pm \pm} &= \frac{1}{U_2 V_2 W_2} \frac{x_1^+ - x_2^+}{x_1^+ - x_2^-}, &\quad D_{\pm \pm} &= U_1 V_1 W_1 \frac{x_1^- - x_2^-}{x_1^+ - x_2^-}, \\
C_{\pm \pm} &= - \frac{\gamma_1}{\gamma_2} \frac{x_2^+ - x_2^-}{x_1^+-x_2^-}, &\quad E_{\pm \pm} &= - \frac{\gamma_2}{\gamma_1} \frac{U_1 V_1 W_1}{U_2 V_2 W_2} \frac{x_1^+ - x_1^-}{x_1^+-x_2^-} , \\
C_{\pm \mp} &= U_1 V_1 W_1 U_2 V_2 W_2 \frac{1-x_1^- x_2^-}{1-x_1^+ x_2^+} , &\quad D_{\pm \mp} &=  1 \mcomma \\
A_{\pm \mp} &= U_2 V_2 W_2 \frac{1-x_1^+ x_2^-}{1-x_1^+ x_2^+}, &\quad E_{\pm \mp} &= U_1 V_1 W_1 \frac{1-x_1^- x_2^+}{1-x_1^+ x_2^+}, \\
B_{\pm \mp} &= i \frac{\gamma_1 \gamma_2}{1-x_1^+ x_2^+}, &\quad F_{\pm \mp}&= - i \frac{U_1 V_1 W_1 U_2 V_2 W_2}{\gamma_1 \gamma_2} \mathrlap{\frac{(x_1^+-x_1^-)(x_2^+-x_2^-)}{1-x_1^+ x_2^+},} \\
\end{aligned}
\end{equation}
\end{fleqn}
and identities
\begin{fleqn}
\begin{equation}
\label{eq:UVW}
U^2 = W^{-2} \frac{x^+ + \xi}{x^-+\xi}= W^2 \frac{x^+}{x^-} \frac{1 + x^- \xi}{1+x^+ \xi} \mcomma \qquad V^2 = W^{-2}  \frac{1 + x^+ \xi}{1+x^- \xi}= W^2 \frac{x^+}{x^-} \frac{x^- + \xi}{x^+ +\xi} \mdot
\end{equation}
\end{fleqn}
Encoded in these relations is the closure condition
\begin{equation}
\label{eq:closure}
\xi^2 (U-U^{-1})^2 -(V-V^{-1})^2+ (1-\xi^2)(W-W^{-1})^2 =0~,
\end{equation}
which can also be written as
\begin{equation}
\label{eq:closure2}
W^{-2} \left(x^+ + \frac{1}{x^+} + \xi + \frac{1}{\xi} \right) = W^{2} \left(x^- + \frac{1}{x^-} + \xi + \frac{1}{\xi} \right)~.
\end{equation}
$\xi$ is a free parameter of the S matrix, which vanishes in the limit $q_L\rightarrow 1$ or $q_R \rightarrow 1$.\footnote{\label{foot:xi} If we denote by $g/2$ the proportionality factor multiplying the two central elements with non-trivial coproduct then
\begin{equation}
\xi^2 = -\frac{(g/2)^2 (q_L-q_L^{-1})(q_R-q_R^{-1})}{1-(g/2)^2(q_L-q_L^{-1})(q_R-q_R^{-1})}~.
\end{equation}}
The relation between $U$, $V$, $W$ and the energy $\omega$, momentum $p$ and charge $\mu$ is given by
\begin{equation}
\label{eq:UVW_rep}
V W^\mu = q_L^{\frac{1}{2}(\omega+\mu)}~, \qquad V W^{-\mu}= q_R^{\frac{1}{2}(\omega-\mu)}~, \qquad U= e^{\frac{ip}{2}}~,
\end{equation}
where $\mu=+1$ for the first representation $\{\ket{\phi_+}, \ket{\psi_+}\}$ and $\mu=-1$ for the second representation $\{\ket{\phi_-}, \ket{\psi_-}\}$.

When writing the action of the S matrix it is implicitly assumed that $U$, $V$, $W$ as well as $x^\pm$ have been evaluated in their respective representation. More precisely, what is meant by $F_{+-}$ for instance is
\begin{equation}
F_{+ -}= - i \frac{U_{1,+} V_{1,+} W_{1,+} U_{2,-} V_{2,-} W_{2,-}}{\gamma_{1,+} \gamma_{2,-}} \frac{(x_{1,+}^+-x_{1,+}^-)(x_{2,-}^+-x_{2,-}^-)}{1-x_{1,+}^+ x_{2,-}^+}~,
\end{equation}
where the $\pm$ subscript refers to the representation of particle 1 and 2.

The parameter $\gamma$ encodes the relative normalization of the fermions $\ket{\psi_\pm}$ with respect to the bosons $\ket{\phi_\pm}$. It is a gauge-like quantity which can be factored out by an appropriate change of basis. It will be convenient to write
\begin{equation}
\label{definition-gamma}
\gamma = \sqrt{i \alpha U V W (x^- - x^+)}~,
\end{equation}
where $\alpha$ is a not-yet specified complex quantity, which can depend on $x^\pm$.
For the special value $\alpha=1$ one exactly recovers the S matrix of \cite{Hoare:2014oua}. For the moment we will keep it unspecified and will choose an expression such that $\alpha \rightarrow 1$ in the undeformed limit, no discontinuities appear in the S matrix due to $\gamma$ and the tree-level expansion matches the perturbative results.

\paragraph{Properties.} This S matrix satisfies the quantum Yang-Baxter equation,
\begin{equation}
\scR_{12} \scR_{13} \scR_{23} =\scR_{23} \scR_{13} \scR_{12}~,
\end{equation}
a version of braiding unitarity
\begin{equation}
\scR_{12} \scR_{21} = \left\{
\begin{aligned} &1 &\qquad &(\pm, \pm) \text{ sector}~, \\
& U_1 V_1 W_1 U_2 V_2 W_2 \frac{1-x_1^- x_2^-}{1-x_1^+ x_2^+} &\qquad &(\pm, \mp) \text{ sector}~,
\end{aligned} \right.
\end{equation}
and crossing relations
\begin{equation}
(\mathcal{C}^{-1} \otimes 1) \scR^{\supertranspose \otimes 1}_{\bar{1}2} (\mathcal C \otimes 1) \scR_{12}  = \left\{
\begin{aligned} &U_1 V_1 W_1 \frac{x_1^- - x_2^+}{x_1^+ - x_2^+} &\qquad &(\pm, \pm) \text{ sector}~, \\
&U_1 V_1 W_1 \frac{1-x_1^- x_2^-}{1-x_1^+ x_2^-} &\qquad &(\pm, \mp) \text{ sector}~.
\end{aligned} \right.
\end{equation}
The $^\supertranspose$ superscript denotes the supertranspose.\footnote{\label{foot:ST} Using the notation $\scR \ket{\Phi_a \Phi_b} = \scR_{ab}^{cd} \ket{\Phi_c \Phi_d}$ with $(\Phi_1,\Phi_2,\Phi_3,\Phi_4)=(\phi_+,\phi_-,\psi_+,\psi_-)$ then $\scR^{\supertranspose \otimes 1} \ket{\Phi_a \Phi_b}=(-1)^{(|a|+1)|c|} \scR_{cb}^{ad} \ket{\Phi_c \Phi_d}$, with $|1|=|2|=0$ (bosonic states) and $|3|=|4|=1$ (fermionic states).} The $\bar{1}$ means that we consider the antipode representation, with $\bar{x}^\pm_{+} = 1/x^{\pm}_{-}$ and $\bar{x}^\pm_{-} = 1/x^{\pm}_{+}$, where again the  $\pm$ subscript denotes the representation.\footnote{We also have $\bar{U}_\pm = U^{-1}_\mp$, $\bar{V}_\pm = V^{-1}_\mp$ and $\bar{W}_\pm = W_\mp$. } The charge conjugation matrix is given by
\begin{equation}
\label{eq:charge_conj}
\mathcal C \ket{\phi_\pm} = \ket{\phi_\mp}~, \qquad \mathcal C \ket{\psi_\pm} = i c_\pm \ket{\psi_\mp}~, \qquad c_\pm =  (\alpha_\pm \bar{\alpha}_\mp)^{-1/2}~.
\end{equation}
Notice that $c_\pm=1$ when $\alpha_\pm=1$.

The S matrix is physically unitary for real momenta and positive energies
\begin{equation}
\scR_{12}^\dagger \scR_{12} = 1~,
\end{equation}
provided that $|\alpha_+|$ and $|\alpha_-|$ are two real numbers (no dependence on $x^\pm$) satisfying $|\alpha_+| |\alpha_-|=1$, and that one imposes the reality conditions $(U^*, V^*, W^*)=(U^{-1}, V, W)$, together with
\begin{equation}
\label{eq:rc1}
 \xi \in i \mathbb{R}~,  \qquad (x^\pm)^* = \frac{x^\mp+\xi}{1+x^\mp \xi}~,
\end{equation}
in the region\footnote{Recall that we restrict to real, positive and non-trivial deformations $q_{L,R} \in \mathbb{R}_{> 0, \neq 1}$.}
\begin{equation}
\label{eq:reg1}
(q_L-q_L^{-1})(q_R-q_R^{-1}) >0 \qquad \Leftrightarrow \qquad (1-q_L)(1-q_R) >0 ~,
\end{equation}
and the reality conditions
\begin{equation}
\label{eq:rc2}
\xi \in (-1,1)~, \qquad (x^\pm)^* = - \frac{x^\mp+\xi}{1+x^\mp \xi}~,
\end{equation}
in the remaining region
\begin{equation}
\label{eq:reg2}
(q_L-q_L^{-1})(q_R-q_R^{-1}) <0 \qquad \Leftrightarrow \qquad  (1-q_L)(1-q_R)<0~.
\end{equation}
For the one-parameter deformation, only the region \eqref{eq:reg1} and reality conditions \eqref{eq:rc1} are possible.

Finally, in the limit $q_{L,R} \rightarrow 1$ with  $\xi \rightarrow 0$ and $\alpha \rightarrow 1$ one recovers the undeformed S matrix of the $\adsThreeT$ superstring with pure Ramond-Ramond fluxes, whose action can directly be obtained from \eqref{eq:Smat} and \eqref{eq:Smat_coeffs} by sending $V \rightarrow 1$, $W \rightarrow 1$ and setting $U=x^+/x^- = e^{ip/2}$. To recover the undeformed version of the closure condition one assumes that $q_{L,R}=e^{-\kappa /g}$ and $\xi =i \kappa$ for small $\kappa$. The parameter $g$ is the one defined in \cref{foot:xi}. Expanding \eqref{eq:closure} and neglecting higher powers of $\kappa$ gives rise to
\begin{equation}
\omega^2 = 1 + 4 g^2 \sin^2 \left(\frac{p}{2}\right)~,
\end{equation}
while \eqref{eq:closure2} gives
\begin{equation}
\left(x^+ + \frac{1}{x^+} \right) - \left(x^- + \frac{1}{x^-} \right) = \frac{2 i}{g}~.
\end{equation}
%{\color{red} The S matrix is not well-defined when only one deformation parameter goes to 1, i.e.~ when $q_L \rightarrow 1, q_R \neq 1$ or $q_R \rightarrow 1, q_L \neq 1$.}

\paragraph{Symmetries.} From the relation \eqref{eq:UVW_rep} it immediately follows that the S matrix has a left-right symmetry
\begin{equation}
\label{eq:LR_sym}
q_L \leftrightarrow q_R ~, \qquad \ket{\phi_\pm} \leftrightarrow \ket{\phi_\mp}~, \qquad \ket{\psi_\pm} \leftrightarrow \ket{\psi_\mp}~.
\end{equation}
This is to be expected since when exchanging $q_L$ and $q_R$ one effectively exchanges the two $\mathfrak{psu}(1|1)$ copies in \eqref{eq:sym_alg_fac}, and this can be reabsorbed into a swapping of the two representations. The central elements do not break this symmetry.

The S matrix is also invariant under\footnote{Under $q_L \rightarrow q_L^{-1}$ we have
\begin{equation*}
U \rightarrow U~, \quad V \rightarrow W^{-\mu}~, \quad W \rightarrow V^{-\mu}~, \quad x^\pm_+ \rightarrow i \sqrt{1-\xi^2} \frac{x^\pm_+}{1+\xi x^\pm_+}~, \quad x^\pm_- \rightarrow -i \frac{\xi+x^\pm_-}{\sqrt{1-\xi^2}}~.
\end{equation*}}
\begin{equation}
\label{eq:Smat_sym_qL}
q_L \rightarrow q_L^{-1}~, \qquad \xi \rightarrow \frac{i \xi}{\sqrt{1-\xi^2}}~, \qquad \alpha_{\pm} \rightarrow \alpha_{\pm} ~,
\end{equation}
and\footnote{Under $q_R \rightarrow q_R^{-1}$ we have
\begin{equation*}
U \rightarrow U~, \quad V \rightarrow W^{\mu}~, \quad W \rightarrow V^{\mu}~,  \quad x^\pm_+ \rightarrow -i \frac{\xi+x^\pm_+}{\sqrt{1-\xi^2}} ~, \quad x^\pm_- \rightarrow i \sqrt{1-\xi^2} \frac{x^\pm_-}{1+\xi x^\pm_-}~.
\end{equation*}}
\begin{equation}
\label{eq:Smat_sym_qR}
q_R \rightarrow q_R^{-1}~, \qquad \xi \rightarrow \frac{i \xi}{\sqrt{1-\xi^2}}~,\qquad \alpha_{\pm} \rightarrow \alpha_{\pm}~.
\end{equation}
These two transformations map the region \eqref{eq:reg1} with reality conditions \eqref{eq:rc1} to the region \eqref{eq:reg2} with reality conditions \eqref{eq:rc2} and vice versa.
We can combine them to deduce that
\begin{equation}
\label{eq:Smat_sym_qLR}
q_L \rightarrow q_L^{-1}~, \qquad q_R \rightarrow q_R^{-1}~, \qquad \xi \rightarrow -\xi~, \qquad \alpha_\pm \rightarrow \alpha_\pm~,
\end{equation}
is another symmetry of the S matrix. Therefore, provided that $\alpha$ remains invariant under inversion of the deformation parameter $q_L$ and/or $q_R$, one has
\begin{equation}
\label{eq:Smat_sym_all}
\scR_{q_L,q_R}= \scR_{q_L^{-1},q_R} =\scR_{q_L,q_R^{-1}}= \scR_{q_L^{-1},q_R^{-1}}~.
\end{equation}
In order to match with the results from perturbation theory and avoid discontinuities in the S matrix we ended up choosing $\alpha$ such that it is not invariant under inversion of the deformation parameter. However, due to the gauge-like nature of $\gamma$, it is important to note that the discrepancy only manifests itself as a one-particle change of basis of the S matrix. The S matrix is thus physically invariant under the above transformations.

\paragraph{Diagonalization.} The S matrix can be diagonalized by means of the nested coordinate Bethe Ansatz. This procedure has been successfully applied to the undeformed S matrix in \cite{Borsato:2012ss}. The results for the deformed case are very similar. We seek to find states $\ket{\Psi}_{\mu_1 \mu_2}$  such that~\footnote{$P^\mup{g}$ is the graded permutation operator, see footnote \ref{definition-graded-embeddings-T}.}
\begin{equation}
\label{eq:diag}
\mathcal R \ket{\Psi}^\mathrm{II}_{\mu_1 \mu_2} = \mathcal R^\mathrm{I,I}_{\mu_1 \mu_2} P^\mup{g}\left(\left.\ket{\Psi}^\mathrm{II}_{\mu_2 \mu_1}\right|_{x_1 \leftrightarrow x_2}\right)~.
\end{equation}
The level-II vacua
\begin{equation}
\label{eq:vac}
\ket{0}^\mathrm{II}_{++} = \ket{\phi_+ \phi_+}, \quad \ket{0}^\mathrm{II}_{--} = \ket{\psi_- \psi_-},\quad \ket{0}^\mathrm{II}_{+-} = \ket{\phi_+ \psi_-},\quad \ket{0}^\mathrm{II}_{-+} = \ket{\psi_- \phi_+},
\end{equation}
trivially satisfy this equation with
\begin{equation}
\scR_{++}^{\mathrm{I,I}}= A_{++} ~, \qquad \scR_{--}^{\mathrm{I,I}}= -F_{--}~, \qquad \scR_{+-}^{\mathrm{I,I}}= C_{+-} ~, \qquad \scR_{-+}^{\mathrm{I,I}}= D_{-+}~.
\end{equation}
The identity \eqref{eq:diag} is also satisfied for the following states with one excitation propagating above the level-II vacuum,
\begin{equation}
\label{eq:prop}
\begin{aligned}
\ket{y}^\mathrm{II}_{++} &= f_+(y,x_1) \ket{\psi_+ \phi_+} + f_+(y,x_2) \mathcal R^\mathrm{II,I}_{++}(y, x_1) \ket{\phi_+ \psi_+}~, \\
\ket{y}^\mathrm{II}_{--} &= f_-(y,x_1) \ket{\phi_- \psi_-} + f_-(y,x_2) \mathcal R^\mathrm{II,I}_{--}(y, x_1) \ket{\psi_- \phi_-}~, \\
\ket{y}^\mathrm{II}_{+-} &= f_+(y,x_1) \ket{\psi_+ \psi_-} + f_-(y,x_2) \mathcal R^\mathrm{II,I}_{-+}(y, x_1) \ket{\phi_+ \phi_-}~, \\
\ket{y}^\mathrm{II}_{-+} &= f_-(y,x_1) \ket{\phi_-\phi_+} + f_+(y,x_2) \mathcal R^\mathrm{II,I}_{+-}(y, x_1) \ket{\psi_- \psi_+}~.
\end{aligned}
\end{equation}
%and
%\begin{equation}
%\begin{aligned}
%\ket{y}^\mathrm{II}_{++, \pi} &= f_+(y,x_2) \ket{ \phi_+ \psi_+} + f_+(y,x_1) \mathcal R^\mathrm{II,I}_{++}(y, x_2) \ket{ \psi_+ \phi_+}~, \\
%\ket{y}^\mathrm{II}_{--, \pi} &= f_-(y,x_2) \ket{\psi_-\phi_- } + f_-(y,x_1) \mathcal R^\mathrm{II,I}_{--}(y, x_2) \ket{ \phi_- \psi_-}~, \\
%\ket{y}^\mathrm{II}_{+-, \pi} &= f_-(y,x_2) \ket{\phi_+ \phi_-} - f_+(y,x_1) \mathcal R^\mathrm{II,I}_{+-}(y, x_2) \ket{\psi_+ \psi_-}~, \\
%\ket{y}^\mathrm{II}_{-+, \pi} &= -f_+(y,x_2) \ket{\psi_-\psi_+} + f_-(y,x_1) \mathcal R^\mathrm{II,I}_{-+}(y, x_2) \ket{\phi_- \phi_+}~.
%\end{aligned}
%\end{equation}
The auxiliary functions are
\begin{equation}
f_+(y,x) = \frac{y \gamma}{h(y)-x^+}~, \qquad f_-(y,x) = \frac{i y}{\gamma}\frac{x^+ - x^-}{1-h(y)x^-}~,
\end{equation}
and the auxiliary S matrix elements read
\begin{equation}
\begin{aligned}
\scR_{++}^{\mathrm{II,I}}(y,x)&= U V W \frac{h(y) -x^-}{h(y)-x^+}~, &\qquad \scR_{--}^{\mathrm{II,I}}(y,x)&= -\frac{1}{U V W} \frac{1-h(y) x^+}{1-h(y) x^-}~, \\
\scR_{-+}^{\mathrm{II,I}}(y,x) &= \scR_{++}^{\mathrm{II,I}}(y,x)~, &\qquad  \scR_{+-}^{\mathrm{II,I}}(y,x) &= \scR_{--}^{\mathrm{II,I}}(y,x)~.
\end{aligned}
\end{equation}
We introduce two auxiliary excitations $y_\pm$ and define the function $h(y_\pm)= (y_\pm)^{\pm 1}$.
These auxiliary excitations then have trivial S matrix $\mathcal R^\mathrm{II,II} (y_{\mu_1,1},y_{\mu_2,2})= 1$.
Due to factorization these results are readily extended to a scattering event involving $M^\mathrm{I}_+$ fundamental representations with $\mu=+1$ and $M^\mathrm{I}_-$ fundamental representations with $\mu = -1$. The Bethe equations for the $M^\mathrm{II}_\pm$ auxiliary excitations $y_\pm$ read
\begin{equation}
\begin{aligned}
1 &= \prod_{j=1}^{M_\pm^\mathrm{I}} \scR_{\pm \pm}^{\mathrm{II,I}}(y_{\pm,k},x_j) \prod_{j=1}^{M_\mp^{\mathrm{I}}} \scR_{\mp \mp}^{\mathrm{II,I}}(y_{\pm ,k},x_j)~, \qquad k=1,\dots, M_\pm^{\mathrm{II}}~.
\end{aligned}
\end{equation}
%Here $J$ is related to the worldsheet length, but only in the full theory.

\paragraph{Dressing factors.} As already mentioned, the symmetries only fix the S matrix up to four dressing factors. It will be convenient to define the ``dressed'' fundamental S matrix $\scS$ through
\begin{equation}
\scS \ket{\Phi_\pm \Phi_\pm}= R_{\pm \pm}  \scR \ket{\Phi_\pm \Phi_\pm}~, \qquad \scS \ket{\Phi_\pm \Phi_\mp}= R_{\pm \mp}  \scR \ket{\Phi_\pm \Phi_\mp}~,
\end{equation}
with $\Phi \in \{\phi,\psi \}$. The four coefficients $R_{++}, R_{+-}, R_{-+}$ and $R_{--}$ are the four dressing factors.\footnote{\label{foot:LR_dressing} In the undeformed case, the discrete left-right symmetry imposes $R_{++}=R_{--}$ as well as $R_{+-}=R_{-+}$, so that there are only two unknown dressing factors~\cite{Borsato:2012ud}. Similarly, in the deformed case, if we assume that the left-right symmetry \eqref{eq:LR_sym} also holds for the dressing factors then we are left with only two unknown quantities $R_{++}$ and $R_{+-}$. The other two dressing factors $R_{--}$ and $R_{-+}$ can be obtained by swapping $q_L \leftrightarrow q_R$.} This dressed S matrix also satisfies the quantum Yang-Baxter equation,
\begin{equation}
\scS_{12} \scS_{13} \scS_{23} =\scS_{23} \scS_{13} \scS_{12}~.
\end{equation}
Moreover, if the dressing factors are pure phases, $|R_{\pm \pm}|=1$ and $|R_{\pm \mp}|=1$, and satisfy
\begin{equation}
\label{eq:cond_dressing}
\begin{aligned}
(R_{\pm \pm})_{12} (R_{\pm \pm})_{21} &=1~, \\
(R_{\pm \mp})_{12} (R_{\mp \pm})_{21} &= \frac{1}{U_{1,\pm} V_{1,\pm} W_{1,\pm} U_{2,\mp} V_{2,\mp} W_{2,\mp}} \frac{1-x_{1,\pm}^+ x_{2,\mp}^+}{1-x_{1,\pm}^- x_{2,\mp}^-} ~, \\
(R_{\mp \pm})_{\bar{1} 2} (R_{\pm \pm})_{12} &= \frac{1}{U_{1,\pm} V_{1,\pm} W_{1,\pm}} \frac{x_{1,\pm}^+ - x_{2,\pm}^+}{x_{1,\pm}^- - x_{2,\pm}^+}~, \\
(R_{\mp \mp})_{\bar{1} 2} (R_{\pm \mp})_{12} &= \frac{1}{U_{1,\pm} V_{1,\pm} W_{1,\pm}} \frac{1-x_{1,\pm}^+ x_{2,\mp}^-}{1-x_{1,\pm}^- x_{2,\mp}^-}~,
\end{aligned}
\end{equation}
then we have braiding unitarity, crossing symmetry and matrix unitarity (upon imposing the reality conditions \eqref{eq:rc1} or \eqref{eq:rc2})
\begin{equation}
\label{eq:Smat_fact_phys_cond}
\scS_{12} \scS_{21} =1~, \qquad (\mathcal{C}^{-1} \otimes 1) \scS^{\supertranspose \otimes 1}_{\bar{1}2} (\mathcal C \otimes 1) \scS_{12}=1 ~, \qquad \scS_{12}^\dagger \scS_{12} =1~.
\end{equation}
Notice that in principle the factorized S matrix does not need to be itself a physical S matrix, but \eqref{eq:Smat_fact_phys_cond} will imply \eqref{eq:Smat_phys_cond}. Solving the equations \eqref{eq:cond_dressing} is an arduous task, that we do not undertake in this paper.

\subsection{Tree-level expansion and matching the perturbative calculations}
Now that we presented the all-loop S matrix based on the conjectured symmetries of the deformed models, we should check that the results are compatible with the perturbative calculations. We assume that in the large tension limit $h \rightarrow \infty$ the deformation parameters behave as
\begin{equation}
q_L = 1-\frac{\kappa_L}{h}+O(h^{-2})~, \qquad q_R = 1- \frac{\kappa_R}{h}+O(h^{-2})~,
\end{equation}
and
\begin{equation}
\xi = \frac{\sqrt{\kappa_-^2-\kappa_+^2}}{\sqrt{1+\kappa_-^2}}+O(h^{-1})~, \qquad \kappa_\pm = \frac{1}{2} \left(\kappa_L \pm \kappa_R \right)~.
\end{equation}
For the unknown parameter $\alpha$, we assume that it has an expansion of the type
\begin{equation}
\label{eq:ltalpha}
\alpha = \varphi^{-2} + O(h^{-2})~,
\end{equation}
with phase $\varphi$ given in \eqref{phase-phi}.
From the relation  $|\alpha| = 1 + O(h^{-2})$ and
\begin{equation}
(1-q_L)(1-q_R) = \frac{\kappa_L \kappa_R}{h^2} +O(h^{-3})= \frac{\kappa_+^2 - \kappa_-^2}{h^2}+O(h^{-3})~,
\end{equation}
it follows that these choices of $\alpha$ and $\xi$ ensure unitarity of the S matrix for all values of (real and unequal) $\kappa_+$ and $\kappa_-$ in the $h \rightarrow \infty$ limit.
Rescaling $p \rightarrow p/h$ and neglecting $O(h^{-2})$ terms gives
\begin{equation}
U = 1+ \frac{i p}{2 h}~, \qquad V=1- \frac{\kappa_+ \omega + \mu \kappa_-}{2 h} ~, \qquad W = 1- \frac{\kappa_+ + \mu \kappa_- \omega}{2 h}~.
\end{equation}
Plugging this into the closure condition \eqref{eq:closure} and choosing the positive energy branch one obtains the dispersion relations in the two different representations,
\begin{equation}
\omega =\mu \kappa_+ \kappa_- + \sqrt{ p^2+m^2}~, \qquad m=\sqrt{(1+\kappa_-^2)(1+\kappa_+^2)}~.
\end{equation}
We then introduce the T matrix as the tree level contribution to the dressed S matrix,
\begin{equation}
\scS=1+\frac{i}{h} \scT+O(h^{-2})~.
\end{equation}
The four dressing factors have an expansion of the type $R_{\mu_1 \mu_2} = 1 + i R_{\mu_1 \mu_2}^{(1)}/h+ \dots$ and will contribute to the diagonal elements of $\scT$. Since we have not solved the conditions \eqref{eq:cond_dressing} we do not have an explicit expression for the dressing factors. We can however find $R_{\mu_1 \mu_2}^{(1)}$ by solving the tree-level version of \eqref{eq:cond_dressing}.
A solution is given by
\begin{equation}
R_{\pm \pm}^{(1)} = -(\mathcal A_{\pm \pm} + \mathcal B_{\pm \pm})~, \qquad R_{\pm \mp}^{(1)} = - \mathcal G_{\pm \mp}~,
\end{equation}
where the coefficients $\mathcal A_{\mu_1 \mu_2}$, $\mathcal B_{\mu_1 \mu_2}$ and $\mathcal G_{\mu_1 \mu_2}$ have been defined in \eqref{T-matrix-coefficients}.\footnote{Notice that 
\begin{equation}
R_{--}^{(1)} = \left. R_{++}^{(1)} \right|_{\kappa_- \rightarrow - \kappa_-}~, \qquad R_{-+}^{(1)} = \left. R_{+-}^{(1)} \right|_{\kappa_- \rightarrow - \kappa_-}~,
\end{equation}
in agreement with footnote \ref{foot:LR_dressing}. }
A tree-level expansion then gives\footnote{We use the same notation as in section \ref{sec:Smat_perturbative}, omitting the two $\pm$ indices for brevity.}
\begin{fleqn}
\begin{equation}
\begin{aligned}
\scT \ket{\phi_\pm \phi_\pm} &= +(\mathcal{A} + \mathcal{B}) \ket{\phi_\pm \phi_\pm} &\qquad  \scT \ket{\psi_\pm \psi_\pm} &= -(\mathcal{A} + \mathcal{B}) \ket{\psi_\pm \psi_\pm} \\
\scT \ket{\phi_\pm \psi_\pm} &= + \mathcal{G} \ket{\phi_\pm \psi_\pm}+ \mathcal H \ket{\psi_\pm \phi_\pm}  &\qquad  \scT \ket{\psi_\pm \phi_\pm} &= - \mathcal{G} \ket{\psi_\pm \phi_\pm}+\mathcal H^* \ket{\phi_\pm \psi_\pm} \\
\scT \ket{\phi_\pm \phi_\mp} &= +\mathcal{A} \ket{\phi_\pm \phi_\mp} +  \mathcal C \ket{\psi_\pm \psi_\mp} &\qquad  \scT \ket{\psi_\pm \psi_\mp} &= - \mathcal{A} \ket{\psi_\pm \psi_\mp} + \mathcal C^* \ket{\phi_\pm \phi_\mp} \\
\scT \ket{\phi_\pm \psi_\mp} &= +\mathcal{G} \ket{\phi_\pm \psi_\mp}  &\qquad  \scT \ket{\psi_\pm \phi_\mp} &= -\mathcal{G} \ket{\psi_\pm \phi_\mp}
\end{aligned}
\end{equation}
\end{fleqn}
where again $\mathcal C_{\mu_1 \mu_2}$ and $\mathcal H_{\mu_1 \mu_2}$ have been defined in \eqref{T-matrix-coefficients}. Labeling the states as in \eqref{particle-index-mapping} this precisely matches the perturbative tree-level expansion \eqref{T-matrix-coefficients}

It would be interesting to find a natural all-loop expression for $\alpha$ which admits \eqref{eq:ltalpha} as large tension expansion. To preserve unitarity we also would like $|\alpha_+|$ and $|\alpha_-|$ to be two real numbers (independent of $x^\pm$) related by $|\alpha_+ ||\alpha_-|=1$. A possibility is to choose
\begin{equation}
\label{eq:all_loop_alpha}
\alpha = \frac{(U-U^{-1} )-( W-W^{-1})}{(U-U^{-1})+(W-W^{-1})}~.
\end{equation}
This is a pure phase $|\alpha|$=1 and is such that $c_\pm =1$ in \eqref{eq:charge_conj}.

%The additional factors of $\varphi$ in the T matrix can be reabsorbed into a field redefinition $\psi_\pm \rightarrow \varphi \psi_\pm$.
The S matrix symmetry transformations \eqref{eq:Smat_sym_qL} and \eqref{eq:Smat_sym_qR} correspond to exchanging $\kappa_+ \leftrightarrow  - \kappa_-$ and $\kappa_+ \leftrightarrow \kappa_-$ respectively, which are symmetries of the functions $\mathcal{A}, \mathcal{B}$ and $\mathcal{G}$. In the functions $\mathcal{C}$ and $\mathcal{H}$ this transformation only affects the phases. Their combination \eqref{eq:Smat_sym_qLR} corresponds to changing the sign of both deformations parameters $\kappa_\pm \rightarrow - \kappa_\pm$, which results into complex conjugation $\varphi \rightarrow \varphi^*$.

\section{\texorpdfstring{Mirror duality of the S matrix for deformed $\adsThreeT$}{Mirror duality of the S matrix for deformed AdS₃⨉S³⨉T⁴}}
\label{sec:mirror_duality}
Superstrings on various $\AdS$ backgrounds (e.g.~$\adsFive$ and $\adsThreeT$) are described by a semi-symmetric space sigma model, with fields that are maps from the two-dimensional worldsheet parameterized by $\tau$ and $\sigma$ to the ten-dimensional target space. The ``mirror'' model is obtained by performing a double Wick rotation and exchanging the time and space coordinates on the worldsheet, $\tau \rightarrow i \tilde{\sigma}$ and $\sigma \rightarrow i \tilde{\tau}$ (we denote the quantities in the mirror model by tildes). In some cases, this mirror theory also describes strings moving on a new, mirror background. The mirror theory can be analyzed in its own right, and one can compute the corresponding perturbative and exact S matrices. For undeformed $\adsFive$ or $\adsThreeT$ superstrings the mirror background and mirror S matrix are different from the original background and S matrix. For one-parameter deformations of $\adsThreeT$ with $q_L = q_R$, it turns out that for the \fermTwo{} background (and only this one), the mirror background is not new, but rather it is related to the original deformed background by an inversion of the deformation parameter $\kappa_+ \rightarrow 1/\kappa_+$. This is the concept of geometric mirror duality. In this section we show that this duality also manifest itself at the level of the exact S matrix. Moreover, due to the invariance (up to a one-particle change of basis) of the S matrix under $q_R \rightarrow q_R^{-1}$, mirror duality will also be present in the other one-parameter limit $q_L = q_R^{-1}$. We however find it unlikely that the notion of mirror duality can be extended to the two-parameter case.

\subsection{Duality of the dispersion relations}
The closure condition \eqref{eq:closure} leads to the dispersion relations
\begin{equation}
\label{eq:mirror_disp}
 \xi^2 \sin^2\left(\frac{p}{2} \right) + \sinh^2\left(\frac{a_+}{2} \omega + \frac{a_-}{2}\mu \right) - (1-\xi^2)\sinh^2 \left(\frac{a_-}{2} \omega + \frac{a_+}{2}\mu \right)=0~,
\end{equation}
where we defined
\begin{equation}
q_L = e^{-a_L}~, \qquad q_R=e^{-a_R}~, \qquad a_\pm = \frac{1}{2} (a_L\pm a_R)~.
\end{equation}
In the special case $a_-=0, a_+ = a$ ($q_L = q_R$) the dispersion relation is invariant under
\begin{equation}
\label{eq:mirror_disp_ap}
p \rightarrow \pm i \omega a~, \qquad \omega \rightarrow \pm \frac{ip}{a}~, \qquad \xi \rightarrow \pm \frac{1}{\xi}~,
\end{equation}
where all choices of sign are allowed. Parametrizing $\xi = i \tan(\theta/2)$, $\theta \in (-\pi, \pi]$ this corresponds to a shift $\theta \rightarrow \pm (\theta + \pi)$. Similarly, in the special case $a_+=0, a_- = a$ ($q_L = q_R^{-1}$) the dispersion relation is invariant under
\begin{equation}
\label{eq:mirror_disp_am}
p \rightarrow \pm i \omega a~, \qquad \omega \rightarrow \pm \frac{ip}{a}~, \qquad \xi \rightarrow \pm \sqrt{1-\xi^2}~,
\end{equation}
which again corresponds to a shift $\theta \rightarrow \pm(\theta + \pi)$ if one defines $\xi = \sin(\theta/2)$.
%The invariance of the dispersion relations under \eqref{eq:mirror_disp_ap} also holds for $\eta$-deformed $\adsFive$.
The invariance under \eqref{eq:mirror_disp_am} is a direct consequence of the invariance under \eqref{eq:mirror_disp_ap} and the invariance (up to a one-particle change of basis, which does not affect the dispersion relation or the spectrum) of the S matrix under inversion of the deformation parameters as stated in \eqref{eq:Smat_sym_all}.
This is a first hint that the S matrix will have ``mirror duality'' when $q_L=q_R$ or $q_L = q_R^{-1}$. Beyond these two cases, both $\sinh^2$ functions in \eqref{eq:mirror_disp} depend on the energy, and while $p \rightarrow -p$ leaves the dispersion relation invariant, this is not the case of the transformation $\omega \rightarrow -\omega$. Hence \eqref{eq:mirror_disp} cannot be invariant under an analytic continuation of the type $p \rightarrow \sharp_1 i \omega$, $\omega \rightarrow \sharp_2 i p$, where $\sharp_{1,2}$ denote energy-independent constants. This hinders the extension of mirror duality to the two-parameter case.\footnote{It was observed in \cite{Hoare:2014oua} that the two-parameter and one-parameter deformations are closely related at the algebraic level. It may thus come as a surprise that the dispersion relations have different behaviors under mirror transformation. This is however to be expected, since the map found in \cite{Hoare:2014oua} also involves the central elements, mixing energy $\omega$, momentum $p$ and charge $\mu$.}

\subsection{Mirror S matrix}
One can study the mirror model in its own right and compute the corresponding S matrix $\tilde{\dsS}(\tilde{x}_1,\tilde{x}_2)$. The symmetries are expected to be the same in the original and mirror theories, up to the identification of the central elements. The mirror S matrix again factorizes, and its matrix part is obtained by analytic continuation
\begin{equation}
\scR(x_1,x_2) \rightarrow \tilde{\scR}(\tilde{x}_1,\tilde{x}_2)~,
\end{equation}
where the arrow denotes the mirror map
\begin{equation}
\label{eq:mirror_pE}
p \rightarrow i \tilde{\omega}~, \qquad \omega \rightarrow i \tilde{p}~.
\end{equation}
In other words, $\tilde{\mathcal R}(\tilde{x}_1,\tilde{x}_2)$ takes the same form as in \eqref{eq:Smat}, but with tilded quantities $\tilde{x}, \tilde{U}, \tilde{V}, \tilde{W}$, and with the identification~
\begin{equation}
\label{eq:UVW_rep_mirror}
\tilde{V} \tilde{W}^\mu = q_L^{\frac{1}{2} (i \tilde{p} + \mu)}~, \qquad \tilde{V}\tilde{W}^{-\mu}= q_R^{\frac{1}{2} (i \tilde{p} - \mu)}~, \qquad \tilde{U} = e^{-\frac{\tilde{\omega}}{2}}~.
\end{equation}
The variables $\tilde{x}^\pm$ satisfy relations similar to \eqref{eq:UVW} and \eqref{eq:closure2}, but where all the quantities now have a tilde, except for $\xi$.\footnote{By definition the mirror transformation preserves the parameters $q_L,q_R$ and $\xi$ of the S matrix.} The mirror S matrix may have a different bound state structure, resulting in new dressing factors, that are not necessarily related to the original dressing factors by a simple analytic continuation.

We want this mirror S matrix to be unitary for real momenta and positive energies. Due to the complex exponents in \eqref{eq:UVW_rep_mirror} the conditions to impose on $\tilde{V}$ and $\tilde{W}$ and ultimately on $\tilde{x}$ are complicated for arbitrary $q_L$ and $q_R$. Here we will focus on the one-parameter case only. Unitarity is ensured if $|\tilde{\alpha}_+|$ and $|\tilde{\alpha}_-|$ are real numbers that do not depend on the momentum which satisfy $|\tilde{\alpha}_+||\tilde{\alpha}_-|=1$ (this in fact follows from the same requirements for the original untilded quantities) and for the reality conditions
\begin{equation}
\xi \in i \mathbb{R}~, \qquad \tilde{U}^* = \tilde{U}~, \qquad \tilde{V}^*=\tilde{V}^{-1}~, \qquad \tilde{W}^* = \tilde{W}~, \qquad (\tilde{x}^\pm)^* = \frac{1+\tilde{x}^\mp \xi}{\tilde{x}^\mp+\xi}~,
\end{equation}
in the case $q_L=q_R$ and
\begin{equation}
\xi \in (-1,+1)~,\qquad \tilde{U}^* = \tilde{U}~, \qquad \tilde{V}^*=\tilde{V}~, \qquad \tilde{W}^* = \tilde{W}^{-1}~, \qquad (\tilde{x}^\pm)^* = \frac{1}{\tilde{x}^\mp}~,
\end{equation}
in the other case $q_L=q_R^{-1}$. These two cases are related through \eqref{eq:Smat_sym_qL} or \eqref{eq:Smat_sym_qR}. Contrary to what is happening for the original S matrix (see \eqref{eq:rc1} and \eqref{eq:rc2}), these conditions break unitarity if one tries to go beyond the one-parameter case.

\subsection{\texorpdfstring{Mirror duality of the S matrix with $q_L=q_R$ or $q_L=q_R^{-1}$}{Mirror duality of the S matrix with qL=qR or qL=qR⁻¹}}
By definition the mirror S matrix will have the same parameters $q_L,q_R, \xi$ as the original S matrix. Mirror duality of the S matrix is then the statement that this mirror S matrix $\tilde{\dsS}(\tilde{x}_1,\tilde{x}_2)$ is equivalent to another S matrix $\dsS(x_1,x_2)$ but now with different parameters $\tilde{q_L}, \tilde{q_R}$ and $\tilde{\xi}$. To show that our exact $q$-deformed S matrix indeed has this property in the one-parameter case we follow \cite{Arutynov:2014ota}, with slight adaptations for the $q_L = q_R^{-1}$ case. We introduce the functions%
\footnote{It is convenient to trade the parameter $\xi$ in favor of $\theta$ to treat the two one-parameter cases in parallel. We define $\xi = i \tan(\theta/2)$ in the case $q_L=q_R$ and $\xi= \sin(\theta)$ in the case $q_L=q_R^{-1}$. In both cases mirror duality will manifest itself through an invariance under a shift $\theta \rightarrow \theta+\pi$.}
\begin{equation}
	\begin{aligned}
	x_s(u; \theta) &= - i \csc(\theta) \left( e^{i u} - \cos(\theta) -(1-e^{i u}) \sqrt{\frac{\cos(u)-\cos(\theta)}{\cos(u)-1}} \right)~, \\
	x_m(u; \theta) &= - i \csc(\theta) \left( e^{i u} - \cos(\theta) +(1+e^{i u}) \sqrt{\frac{\cos(u)-\cos(\theta)}{\cos(u)+1}} \right)~, \\
	y_{s,m}(u; \theta) &= i \cos(\frac{\theta}{2}) x_{s,m}(u; \theta) - \sin(\frac{\theta}{2})~, \\
	\frac{1}{z_{s,m}(u; \theta)} &= i \cos(\frac{\theta}{2}) \frac{1}{x_{s,m}(u;\theta)} - \sin(\frac{\theta}{2})~.
	\end{aligned}
\end{equation}
In the case $q_L=q_R = e^{-a}$ we define $x^\pm = x_s(u\pm i a; \theta)$ and $\tilde{x}^\pm = x_m(u\pm i a; \theta)$, so that the equation \eqref{eq:closure2} as well as its tilded counterpart are satisfied. For $u \in \mathbb{R}$ one then has the desired reality conditions
\begin{equation}
(x^\pm)^* = \frac{x^\mp+\xi}{1+x^\mp \xi} ~, \qquad (\tilde{x}^\pm)^* = \frac{1+\tilde{x}^\mp \xi}{\tilde{x}^\mp+\xi}~.
\end{equation}
In the case $q_L=q_R^{-1}= e^{-a}$ we define $x^\pm_+ = y_s(u + i a; \theta)$, $x^\pm_- = z_s(u + i a; \theta)$ and $\tilde{x}^\pm_+ = y_m(u\pm i a; \theta)$, $\tilde{x}^\pm_- = z_m(u\pm i a; \theta)$, so that the equation \eqref{eq:closure2} as well as its tilded counterpart are again satisfied for both representations. These functions satisfy the desired reality conditions for $u \in \mathbb{R}$, namely
\begin{equation}
(x^\pm)^* = -\frac{x^\mp+\xi}{1+x^\mp \xi} ~, \qquad (\tilde{x}^\pm)^* = \frac{1}{\tilde{x}^\mp}~.
\end{equation}
Moreover, we have the relation
\begin{equation}
x_s(u+\pi; \theta+\pi) = x_m(u; \theta)~,
\end{equation}
which translates to
\begin{equation}
	\begin{aligned}
	\sin (\frac{\theta}{2}) y_s(u + \pi; \theta+\pi) + \frac{1}{2} &= \cos(\frac{\theta}{2}) y_m(u; \theta) - \frac{1}{2}~, \\
	\cos (\frac{\theta}{2}) \frac{1}{z_s(u + \pi; \theta+\pi)} + \frac{1}{2} &= -\sin(\frac{\theta}{2}) \frac{1}{z_m(u; \theta)} - \frac{1}{2}~.
	\end{aligned}
\end{equation}
For both cases this relation implies that
\begin{equation}
	\begin{aligned}
	U(u+\pi; \theta+\pi) &=  \tilde{V}(u; \theta) &\qquad &\Leftrightarrow &\qquad p(u+\pi; \theta+\pi) &= -a \tilde{p}(u; \theta)~, \\
	V(u+\pi; \theta+\pi) &= \tilde{U}(u; \theta) &\qquad &\Leftrightarrow &\qquad \omega(u+\pi; \theta+\pi) &= \frac{1}{a}\tilde{\omega}(u; \theta)~.
	\end{aligned}
\end{equation}
Plugging into \eqref{eq:mirror_pE} this explains the invariance of the dispersion relation under \eqref{eq:mirror_disp_ap} and \eqref{eq:mirror_disp_am}.

At the level of the factorized S matrix, mirror duality is realized if
\begin{equation}
\label{eq:mirror_duality}
D \scS(p_1(u+\pi; \theta+\pi),p_2(u+\pi; \theta+\pi); \theta+\pi) D^{-1}= \tilde{\scS}(-\tilde{p}_1(u; \theta),-\tilde{p}_2(u; \theta); \theta)~,
\end{equation}
where $D$ is the matrix realization of a permutation exchanging bosons and fermions. This additional operator needs to be introduced to account for the fact that under a mirror transformation, the $\AdS$ and sphere coordinates are exchanged, which results into a swapping $\ket{\phi} \leftrightarrow \ket{\psi}$, as follows from \eqref{particle-index-mapping}. It is convenient to split \eqref{eq:mirror_duality} into relations for the matrix part and the dressing phases of the S matrix.

The matrix part of the S matrix, with the choice \eqref{eq:all_loop_alpha} for $\alpha$, satisfies a relation similar to the one highlighted in \cite{Arutynov:2014ota}, namely
\begin{align}
\label{eq:mirrorAdS3}
D \scR(p_1,p_2) D^{-1} = \left\{ \begin{aligned}
&A_{\pm \pm}(p_1,p_2) \scR(-p_1,-p_2) &\qquad &(\pm,\pm) \text{ sector}~, \\
&C_{\pm \mp}(p_1,p_2) \scR(-p_1,-p_2) &\qquad &(\pm,\mp) \text{ sector}~,
\end{aligned}
\right.
\end{align}
with
\begin{equation}
\label{eq:defD}
\begin{aligned}
D \ket{\phi_a \phi_b} &= - \ket{\psi_a \psi_b}~,  \\
D \ket{\phi_a \psi_b} &= - \ket{\psi_a \phi_b}~,  \\
D \ket{\psi_a \phi_b} &= + \ket{\phi_a \psi_b}~,\\
D \ket{\psi_a \psi_b} &= - \ket{\phi_a \phi_b}~.
\end{aligned}
\end{equation}
This identity is in fact satisfied for all $q_L$ and $q_R$. If one restricts to the $(\pm, \pm)$ sectors then the first line of \eqref{eq:mirrorAdS3} is also satisfied for $D = (-1)^F M \otimes M$, $F$ being the fermion number operator and $M$ the matrix representation of the permutation $(3,4,1,2)$, i.e.
\begin{equation}
M \ket{\phi_\pm} = \ket{\psi_\pm}~, \qquad M \ket{\psi_\pm} = \ket{\phi_\pm}~.
\end{equation}
This is precisely the equation highlighted in \cite{Arutynov:2014ota} for the distinguished $\mathfrak{su}(2|2)$ invariant S matrix of $\adsFive$.\footnote{The equation looks different from the one in \cite{Arutynov:2014ota}, but this is only due to the fact that the authors use the graded S matrix. Our S matrix is related to the one in the mentioned paper through an additional minus sign when the two outgoing particles are fermions, as well as an overall factor.} The relation does however not hold in the $(\pm, \mp)$ sectors.\footnote{To have $D= (-1)^F M \otimes M$ for all sectors the permutation matrix should be modified to $M \ket{\phi_\pm} = \ket{\psi_\pm}$, $M \ket{\psi_\pm} = i \ket{\phi_\pm}$.}

Consequently, if the dressing factors satisfy
\begin{equation}
	\begin{aligned}
	\frac{R_{\pm \pm}(p_1(u+\pi; \theta+\pi),p_2(u+\pi; \theta+\pi); \theta+\pi)}{\tilde{R}_{\pm \pm}(-\tilde{p}_1(u; \theta),-\tilde{p}_2(u; \theta); \theta)} &= \frac{1}{A_{\pm \pm}(p_1,p_2)}~, \\
	\frac{R_{\pm \mp}(p_1(u+\pi; \theta+\pi),p_2(u+\pi; \theta+\pi); \theta+\pi)}{\tilde{R}_{\pm \mp}(-\tilde{p}_1(u; \theta),-\tilde{p}_2(u; \theta); \theta)} &=  \frac{1}{C_{\pm \mp}(p_1,p_2)}~,
	\end{aligned}
\end{equation}
then \eqref{eq:mirror_duality} is satisfied. From this analysis we conclude that, provided the conditions on the dressing factors are satisfied, given a $q$-deformed S matrix with parameter $\theta$ (or $\xi$), its mirror S matrix (which by definition also has parameter $\theta$) is then equivalent to another $q$-deformed S matrix but with parameter $\theta+\pi$ (or $1/\xi$ in the $q_L=q_R$ case, $\sqrt{1-\xi^2}$ in the $q_L = q_R^{-1}$ case).

\section{\texorpdfstring{Mirror duality of the fermionic S matrix for $\adsFive$}{Mirror duality of the fermionic S matrix for AdS₅⨉S⁵}}
\label{sec:mirror_duality_AdS5}
To conclude the analysis of mirror duality in $\eta$-deformed theories, we consider the case of the $\eta$-deformed $\adsFive$ superstring based on the distinguished and fully fermionic Dynkin diagrams of $\mathfrak{psu}(2,2|4)$. The background associated to the distinguished deformation can be found in \cite{Arutyunov:2015qva} and does not exhibit geometric mirror duality. The distinguished S matrix~\cite{Beisert:2008tw} was however shown to have spectrum mirror duality in \cite{Arutynov:2014ota}. As far as the fermionic deformation is concerned, the background obtained in \cite{Hoare:2018ngg} is also not geometric mirror dual. Its corresponding factorized S matrix~\cite{Seibold:2020ywq} reads
\begin{fleqn}
\begin{equation}
\resizeToFitPageMath{
\begin{aligned}
\scR \ket{\phi_a \phi_b} &= A \ket{\phi_a \phi_b}, \qquad \qquad \qquad \qquad \qquad \qquad  \scR \ket{\psi_a \psi_b} = -D \ket{\psi_a \psi_b} , \\
\scR \ket{\phi_1 \phi_2} &= \frac{A-B}{q+q^{-1}} \ket{\phi_1 \phi_2} + \varphi_{12} \frac{\hat{a}_1}{\hat{a}_2} \frac{q A + q^{-1} B}{q+q^{-1}} \ket{\phi_2 \phi_1} + \frac{g_2}{f_1}  \frac{ \hat{b}_1}{ \hat{a}_2}  \frac{q C}{q+q^{-1}}\ket{\psi_3 \psi_4} - \frac{\hat{b}_2}{\hat{a}_2} \frac{q^2 C}{q+q^{-1}} \ket{\psi_4 \psi_3}, \\
\scR \ket{\phi_2 \phi_1} &= \hat{\varphi}_{12} \frac{\hat{a}_2}{\hat{a}_1}\frac{q^{-1} A + q B}{q+q^{-1}} \ket{\phi_1 \phi_2} +  \frac{A-B}{q+q^{-1}}  \ket{\phi_2 \phi_1} -  \frac{ \hat{b}_1}{ \hat{a}_1}  \frac{q^2 C}{q+q^{-1}}\ket{\psi_3 \psi_4} + \frac{g_1}{f_2}\frac{\hat{b}_2}{\hat{a}_1} \frac{q C}{q+q^{-1}} \ket{\psi_4 \psi_3}, \\
\scR \ket{\psi_3 \psi_4} &= -\frac{D-E}{q+q^{-1}} \ket{\psi_3 \psi_4} - \varphi_{21} \frac{\hat{b}_2}{\hat{b}_1} \frac{q D + q^{-1} E}{q+q^{-1}} \ket{\psi_4 \psi_3} - \frac{f_1}{g_2}  \frac{ \hat{a}_2}{ \hat{b}_1}  \frac{q^{-1} F}{q+q^{-1}}\ket{\phi_1 \phi_2} +\frac{\hat{a}_1}{\hat{b}_1} \frac{q^{-2} F}{q+q^{-1}} \ket{\phi_2 \phi_1}, \\
\scR \ket{\psi_4 \psi_3} &= - \hat{\varphi}_{21} \frac{\hat{b}_1}{\hat{b}_2} \frac{q^{-1} D + q E}{q+q^{-1}} \ket{\psi_3 \psi_4}  -\frac{D-E}{q+q^{-1}} \ket{\psi_4 \psi_3} + \frac{ \hat{a}_2}{ \hat{b}_2}  \frac{q^{-2} F}{q+q^{-1}}\ket{\phi_1 \phi_2} - \frac{f_2}{g_1}\frac{\hat{a}_1}{\hat{b}_2} \frac{q^{-1} F}{q+q^{-1}} \ket{\phi_2 \phi_1}, \\
\scR \ket{\phi_a \psi_b} &= G \ket{\phi_a \psi_b} + H_{ab} \ket{\psi_b \phi_a}, \qquad \qquad \qquad \scR \ket{\psi_a \phi_b} = L  \ket{\psi_a \phi_b} + K_{ab} \ket{\phi_b \psi_a},
\end{aligned}
}
\end{equation}
\end{fleqn}
where
\begin{equation}
\begin{aligned}
\begin{aligned}
H_{13} &= \frac{f_2}{f_1} \frac{\hat{b}_1}{\hat{b}_2} H~, &\qquad H_{14}&= H~, &\qquad H_{23}&= \frac{\hat{a}_2}{\hat{a}_1} \frac{\hat{b}_1}{\hat{b}_2} H~, &\qquad H_{24}&= \frac{g_1}{g_2} \frac{\hat{a}_2}{\hat{a}_1} H~, \\
K_{13} &= \frac{f_1}{f_2} \frac{\hat{b}_2}{\hat{b}_1} K~, &\qquad K_{14}&= K~, &\qquad K_{23}&= \frac{\hat{a}_1}{\hat{a}_2} \frac{\hat{b}_2}{\hat{b}_1} K~, &\qquad K_{24}&= \frac{g_2}{g_1} \frac{\hat{a}_1}{\hat{a}_2} K~,
\end{aligned}
\\
\varphi_{12}=\frac{q^{1/2} f_1 x_1^-+q^{-2 C_1-1/2} g_1 x_2^+}{q^{1/2} f_1 x_1^-+q^{-2 C_2-1/2} g_1 x_2^+}~, \qquad \hat{\varphi}_{12}=\frac{q^{-1/2} g_2 x_1^- +q^{-2 C_2+1/2} f_2 x_2^+}{q^{-1/2} g_2 x_1^- +q^{-2 C_1+1/2} f_2 x_2^+}~,
\end{aligned}
\end{equation}
and
\begin{equation}
f_j = 1+\xi/x_j^-~, \qquad g_j = 1+ \xi x_j^+~.
\end{equation}
The ten coefficients $A,B,\dots,L$ can be found in \cite{Seibold:2020ywq}. We take\footnote{With respect to \cite{Seibold:2020ywq} there is an additional factor of $i$ in $\gamma$. This is to match with the convention used in \cite{Arutynov:2014ota}. It can be reabsorbed into a rescaling of the fermions.}
\begin{equation}
\gamma = q^{-3/2} \sqrt{i q^{1/2} U V (x^- - x^+)}~,
\end{equation}
together with the reality conditions
\begin{equation}
\hat{a}_j = q^{C_j-1/2}~, \qquad \hat{b}_j = q^{C_j+1/2}~.
\end{equation}

This fermionic S matrix satisfies the same relation as its distinguished counterpart, namely
\begin{equation}
D \scR(p_1,p_2) D^{-1} = A \scR(-p_1,-p_2)~,
\end{equation}
where $D= M \otimes (-1)^{F} M$. Therefore also the fermionic S matrix will have spectrum mirror duality.

\section{Conclusions}

In this paper we computed the tree level S matrices for two parameter $\eta$ deformations of $\adsThreeT$ for various choices of Dynkin diagram underlying the deformation. Given judicious choices of mode expansion summarized in table \ref{table-phases-mode-expansion}, these tree level S matrices all agree with the perturbative expansion of the exact two parameter $q$-deformed S matrix. Next, we investigated the mirror duality properties of this exact $S$ matrix in the one-parameter deformation limit, as well as the exact $S$ matrix for fermionic $\eta$-deformed $\adsFive$. We showed that both of these exact S matrices are compatible with mirror duality, where in the $\adsThreeT$ case we assume certain properties of the thus far undetermined $q$-deformed dressing phases.

There are various open questions associated to these models and S matrices. In terms of deformations of $\adsThreeT$, it would be interesting to determine the deformed dressing phases for the two parameter deformed S matrix. Moreover, particularly for the two parameter deformation it would be insightful to investigate the mirror theory. Based on e.g.~the more involved structure of the dispersion relation \eqref{eq:mirror_disp}, we expect this to be quite different from the single parameter case. It would also be good to understand the full background of the three parameter deformation \cite{Delduc:2018xug}, and determine its S matrix and the effect of the various deformations on the spectrum of this theory. Then, coming back to $\adsFive$, here the relation between the mirror $\adsFive$ background and inhomogeneous deformations of $\adsFive$, is still a bit of a mystery that deserves further investigation.\footnote{Mirror $\adsFive$ by definition corresponds to an integrable sigma model, it is one-loop Weyl invariant, and it is spectrally equivalent to a maximal deformation limit of $\eta$ deformed $\adsFive$, while at the same time, geometrically it differs from the maximal deformation limit of all known inhomogeneous deformations of $\adsFive$.} Moreover, it would be interesting to investigate whether the Bethe ansatz equations and spectrum for the fermionic deformation of $\adsFive$ differ from those of the distinguished deformation, and if so, to determine the effect of the change of Dynkin diagram on the spectrum.
In general, it would be nice to extend the perturbative computation of these deformed $\textup{AdS}_n \times \textup{S}^n$ S matrices to loop level.\footnote{For the distinguished case the one-loop S matrix has been studied using unitarity techniques in \cite{Engelund:2014pla}.} Moreover, it would be interesting to investigate a possible algebraic interpretation of the differences in the mode expansion (one particle change of basis) of table \ref{table-phases-mode-expansion}, related to the choice of Dynkin diagram determining the deformation, as well as a presumably similar interpretation of the results of \cite{Hoare:2016ibq} relating to different implementations of the real form of the algebra. Finally of course it would be exciting to understand whether these deformations can be implemented in the field theories dual to the undeformed $\adsThreeT$ and $\adsFive$ strings.

\section*{Acknowledgments}

We thank Riccardo Borsato, Ben Hoare and Alessandro Sfondrini for discussions, and Riccardo Borsato, Ben Hoare and Arkady Tseytlin for comments on the draft of this paper. The work of FS is supported by the Swiss National Science Foundation via the Early Postdoc.Mobility fellowship ``q-deforming AdS/CFT''. The work of ST and YZ is supported by the German Research Foundation (DFG) via the Emmy Noether program ``Exact Results in Extended Holography''. ST is supported by LT.

\appendix

\section{Generalized supergravity backgrounds}
\label{app:generaliedsugrabackgrounds}

Here we provide the fluxes $\mathcal{F}$ for the generalized supergravity backgrounds corresponding to deformations based on the distinguished and \xox{} Dynkin diagrams of $\mathfrak{psu}(1,1|2)$, as derived in \cite{Seibold:2019dvf}.\footnote{We take this opportunity to correct a typo in the five-forms $\mathcal F_5$ of \cite{Seibold:2019dvf}. The term $J_2 \wedge J_2$ should be replaced by $\text{d} x^1 \wedge \text{d} x^2 \wedge \text{d} x^3 \wedge \text{d} x^4 = - \frac{1}{2} J_2 \wedge J_2$. } The first distinguished background, \distOne{}, has
\begin{equation}
\begin{aligned}
\label{eq:distinguishedbackground1}
\mathcal F_1 &=  N \, \hat{\mathcal F}_1 ~, \\
\mathcal F_3 &=  N \left(\hat{\mathcal F}_3 + \frac{2 \kappa_-}{1-\kappa_-^2} \hat{\mathcal F}_1 \wedge J_2\right) ~,\\
\mathcal F_5 &= N \left( - \frac{1}{2} (1+\star) \hat{\mathcal F}_1 \wedge J_2 \wedge J_2+ \frac{2 \kappa_-}{1-\kappa_-^2} \hat{\mathcal F}_3 \wedge J_2 \right)  ~,
\end{aligned}
\end{equation}
where $J_2$ is given in equation \eqref{eq:torusKahlerform},
\begin{equation}
N = 2 \sqrt{\frac{1+\kappa_+^2}{1+\kappa_-^2}} \frac{1-\kappa_-^2}{\sqrt{F(\rho)\tilde{F}(r)}}~,
\end{equation}
and
\begin{equation}
\begin{aligned}
\label{eq:distinguishedauxiliaryforms}
\hat{\mathcal F}_1 &= \kappa_- \left[ (1+\rho^2) \dd t + (1-r^2) \dd \varphi \right] + \kappa_+ \left[-\rho^2 \dd \psi + r^2 \dd \phi \right] ~, \\
\hat{\mathcal F}_3 &= \frac{1}{F(\rho)} \big[ \rho \, \dd t \wedge \dd \rho \wedge \dd \psi - \kappa_+^2 \rho r^2 \,\dd t \wedge \dd \rho \wedge \dd \phi - \kappa_-^2 \rho (1-r^2) \,\dd \rho \wedge \dd \psi \wedge \dd \varphi  \\
&\qquad- \kappa_+ \kappa_- \rho (1-r^2) \,\dd t \wedge \dd \rho \wedge \dd \varphi - \kappa_+ \kappa_- \rho r^2 \,\dd \rho \wedge \dd \psi \wedge \dd \phi \big] \\
& + \frac{1}{\tilde{F}(r)} \big[ r \, \dd \varphi \wedge \dd r \wedge \dd \phi + \kappa_+^2 \rho^2 r \, \dd \psi \wedge \dd \varphi \wedge \dd r - \kappa_-^2 (1+\rho^2) r \, \dd t \wedge \dd r \wedge \dd \phi \\
&\qquad - \kappa_+ \kappa_- (1+\rho^2) r \, \dd t \wedge \dd \varphi \wedge \dd r + \kappa_+ \kappa_- \rho^2 r \, \dd \psi \wedge \dd r \wedge \dd \phi \big] ~.
\end{aligned}
\end{equation}
The second distinguished background, \distTwo{}, has
\begin{equation}
\begin{aligned}
\label{eq:distinguishedbackground2}
\mathcal F_1 &=  M \,\hat{\mathcal F}_1 ~, \\
\mathcal F_3 &=  M \left(\hat{\mathcal F}_3 + \frac{2 \kappa_+}{1-\kappa_+^2} \hat{\mathcal F}_1 \wedge J_2\right) ~,\\
\mathcal F_5 &= M \left( - \frac{1}{2}(1+\star) \hat{\mathcal F}_1 \wedge J_2 \wedge J_2+ \frac{2 \kappa_+}{1-\kappa_+^2} \hat{\mathcal F}_3 \wedge J_2 \right)  ~,
\end{aligned}
\end{equation}
where
\begin{equation}
M = 2 \sqrt{\frac{1+\kappa_-^2}{1+\kappa_+^2}} \frac{1-\kappa_+^2}{\sqrt{F(\rho)\tilde{F}(r)}}~.
\end{equation}
Note that the dependence on $\kappa_+$ and $\kappa_-$ is exchanged between equations \eqref{eq:distinguishedbackground1} and \eqref{eq:distinguishedbackground2}, except that it remains the same in equations \eqref{eq:distinguishedauxiliaryforms}.

Finally, the first \xox{} deformation, \xoxOne{}, corresponds to
\begin{equation}
\begin{aligned}
\label{eq:backoxooxoTsT}
\mathcal F_1 &=  L \, \hat{\mathcal F}_1 ~, \\
\mathcal F_3 &=  L \, \hat{\mathcal F}_3 ~,\\
\mathcal F_5 &=  - \frac{L}{2}  (1+\star) \hat{\mathcal F}_1 \wedge J_2 \wedge J_2~,
\end{aligned}
\end{equation}
with
\begin{equation}
L = 2 \frac{\sqrt{(1+\kappa_-^2)(1+\kappa_+^2)}}{\sqrt{F(\rho)\tilde{F}(r)}}~,
\end{equation}
and a change of sign of $t$ and $\psi$ in the expressions for the $\hat{\mathcal{F}}$.
The second \xox{} deformation, \xoxTwo{}, is obtained from \xoxOne{} through the map
\begin{equation}
\rho \to \frac{i \sqrt{1+\kappa_-^2} \sqrt{1+\rho ^2}}{\sqrt{F(\rho)}}~,\qquad r\to \frac{\sqrt{1+\kappa_-^2} \sqrt{1-r^2}}{\sqrt{\tilde{F}(r)}}~,\qquad t\leftrightarrow \psi ~, \qquad \varphi \leftrightarrow \phi~.
\end{equation}

\section{Feynman diagrammatics}
\label{feynman-diagrammatics}
We calculated the entries of the perturbative T matrix with the use of Feynman diagrams.
The implementation uses \emph{Mathematica} together with the packages \emph{FeynRules} \cite{Alloul:2013bka} and \emph{FeynArts} \cite{Hahn:2000kx}.
The code is directly derived from our previous work on $\eta$-deformed $\adsFive$ \cite{Seibold:2020ywq} and we refer the reader to its appendix B for a detailed account of the implementation.
All modifications performed for the present case are described in the following.

The $\adsThree$ model only has two massive complex bosons and four massive complex fermions.
We regard the former as four real bosons.
The latter become real on shell, and therefore the treatment of \cite{Seibold:2020ywq} still applies.
Due to the different quadratic Lagrangian \eqref{quadratic-lagrangian} and mode expansions \eqref{mode-expansion} in comparison to the $\adsFive$ model, the prefactors and fermionic wave functions change:
The prefactors become $\frac{1}{2 \sqrt{\omegaNormalization_p}}$ and $\frac{1}{\sqrt{2 \omegaNormalization_p}}$ for bosons and fermions respectively.
The bosonic wave functions are $\operatorPhaseYPlusMinus$ and $\operatorPhaseZPlusMinus$ for incoming $Y$ and $Z$ particles respectively and $\conjugated{\operatorPhaseYPlusMinus}$ and $\conjugated{\operatorPhaseZPlusMinus}$ for outgoing ones.
The fermionic wave functions are (in the notation of figure~2.3 of \cite{Denner:1992vza})
\begin{equation}
	\begin{aligned}
		u_{\zeta_L} &= - \fermionPhase \, f_{+p} \, \operatorPhaseZetaPlus
		\qquad \qquad &
		u_{\zeta_R} &= - \fermionPhase \, f_{-p} \, \operatorPhaseZetaPlus
		\\
		\bar{u}_{\zeta_L} &= \fermionPhase \, f_{+p} \, \conjugated{\operatorPhaseZetaMinus}
		&
		\bar{u}_{\zeta_R} &= - \fermionPhase \, f_{-p} \, \conjugated{\operatorPhaseZetaMinus}
		\\
		v_{\zeta_L} &= - \conjugated{\fermionPhase} f_{+p} \, \conjugated{\operatorPhaseZetaPlus}
		&
		v_{\zeta_R} &= - \conjugated{\fermionPhase} f_{-p} \, \conjugated{\operatorPhaseZetaPlus}
		\\
		\bar{v}_{\zeta_L} &= \conjugated{\fermionPhase} f_{+p} \, \operatorPhaseZetaMinus
		&
		\bar{v}_{\zeta_R} &= - \conjugated{\fermionPhase} f_{-p} \, \operatorPhaseZetaMinus
		\\[10pt]
		u_{\chi_L} &= \fermionPhase \, f_{+p} \, \operatorPhaseChiPlus
		&
		u_{\chi_R} &= \fermionPhase \, f_{-p} \, \operatorPhaseChiPlus
		\\
		\bar{u}_{\chi_L} &= - \fermionPhase \, f_{+p} \, \conjugated{\operatorPhaseChiMinus}
		&
		\bar{u}_{\chi_R} &= \fermionPhase \, f_{-p} \, \conjugated{\operatorPhaseChiMinus}
		\\
		v_{\chi_L} &= \conjugated{\fermionPhase} f_{+p} \, \conjugated{\operatorPhaseChiPlus}
		&
		v_{\chi_R} &= \conjugated{\fermionPhase} f_{-p} \, \conjugated{\operatorPhaseChiPlus}
		\\
		\bar{v}_{\chi_L} &= - \conjugated{\fermionPhase} f_{+p} \, \operatorPhaseChiMinus
		&
		\bar{v}_{\chi_R} &= \conjugated{\fermionPhase} f_{-p} \, \operatorPhaseChiMinus
	\end{aligned}
\end{equation}
where $\fermionPhase$ and the various $\operatorPhase_x$ are defined in \cref{table-phases-mode-expansion}.

\newcommand{\FeynAmplitude}{\mathcal{M}}
The dispersion \eqref{perturbative-dispersion-relation} differs from the $\adsFive$ case, and in particular depends on the particle species.
This changes the relation between $\dsT$ and the modified Feynman amplitudes $\FeynAmplitude$ to
\begin{align*}
	\dsT(p_1, p_2) & =
		\int \dd{k_1} \dd{k_2} \delta(p_1 + p_2 - k_1 - k_2) \delta(\omega_{p_1} + \omega_{p_2} - \omega_{k_1} - \omega_{k_2}) \; \FeynAmplitude(p_1, p_2, k_1, k_2)
	\\ & =
		\frac{\omegaNormalization_{p_1} \omegaNormalization_{p_2}}{\abs{p_1 \omegaNormalization_{p_2} - p_2 \omegaNormalization_{p_1}}} (\FeynAmplitude(p_1, p_2, p_1, p_2) + \FeynAmplitude(p_1, p_2, p_2, p_1))
	\mdot
	\numberthis
\end{align*}
Here $p_i$ and $k_i$ denote the momenta of and simultaneously label the incoming and outgoing particles.
Due to energy and momentum conservation the outgoing momenta are restricted to take on the same values as the incoming momenta.
To follow the existing literature we assume that $p_1 > p_2$.
We want to highlight that even though the energies $\omega_i = \mu_i \kappa_+ \kappa_- + \omegaNormalization_i$ contain a constant shift dependent on the \enquote{charge} $\mu_i$ of the particle species, these contributions cancel in the energy-conservation delta function and only the $\omegaNormalization_i$ contribute to the final kinematic factor of the second line.
This is due to the observation that only amplitudes for which $\mu_{p_1} + \mu_{p_2} = \mu_{k_1} + \mu_{k_2}$ holds are non-zero, i.e.\ $\mu$ is a conserved.

\pdfbookmark[section]{\refname}{bib} % add bibliography to pdf bookmarks
\begin{sloppypar} % sloppy line breaking in bibliography
\bibliographystyle{nb}
\bibliography{bibliography}
\end{sloppypar}

\end{document}